\documentclass[reprint,
 superscriptaddress,
amsmath,amssymb,
prb,
]{revtex4-2}

\usepackage{graphicx}
\usepackage{dcolumn}
\usepackage{bm}
\usepackage{xcolor}

\begin{document}
\preprint{APS/123-QED}
\title{Spectroscopic evidence for topological band structure in FeTe$_{0.55}$Se$_{0.45}$}

\author{Y.-F. Li}
\thanks{These authors contributed equally.}
 \affiliation{Stanford Institute for Materials and Energy Sciences, SLAC National Accelerator Laboratory, Menlo Park, 94025, CA, USA}
\affiliation{Department of Applied Physics and Physics, Stanford University, Stanford, 94305, CA, USA}
\affiliation{Geballe Laboratory for Advanced Materials, Stanford University, Stanford, 94305, CA, USA}

\author{S.-D. Chen}
\thanks{These authors contributed equally.}
\affiliation{Stanford Institute for Materials and Energy Sciences, SLAC National Accelerator Laboratory, Menlo Park, 94025, CA, USA}
\affiliation{Department of Applied Physics and Physics, Stanford University, Stanford, 94305, CA, USA}
\affiliation{Geballe Laboratory for Advanced Materials, Stanford University, Stanford, 94305, CA, USA}
\affiliation{Department of Physics, University of California, Berkeley, California 94720, USA}

\author{M. García-Díez}
\affiliation{Donostia International Physics Center, 20018 Donostia-San Sebastián, Spain}
\affiliation{Physics Department, University of the Basque Country (UPV/EHU), Bilbao, Spain}

\author{M. I. Iraola}
\affiliation{Donostia International Physics Center, 20018 Donostia-San Sebastián, Spain}

\author{H. Pfau}
 \affiliation{Stanford Institute for Materials and Energy Sciences, SLAC National Accelerator Laboratory, Menlo Park, 94025, CA, USA}
\affiliation{Advanced Light Source, Lawrence Berkeley National Laboratory, Berkeley, 94720, CA, USA}
\affiliation{Department of Physics, Pennsylvania State University, University Park, 16802, PA, USA}

\author{Y.-L. Zhu}\author{Z.-Q. Mao}
\affiliation{Department of Physics, Pennsylvania State University, University Park, 16802, PA, USA}

\author{T. Chen}\author{M. Yi}\author{P.-C. Dai}
\affiliation{Department of Physics and Astronomy, Rice University, Houston, 77005, TX, USA}

\author{J.A. Sobota}
 \affiliation{Stanford Institute for Materials and Energy Sciences, SLAC National Accelerator Laboratory, Menlo Park, 94025, CA, USA}

\author{M. Hashimoto}
\affiliation{Stanford Synchrotron Radiation Lightsource, SLAC National Accelerator Laboratory, Menlo Park, 94025, CA, USA}

\author{M. G. Vergniory}
\email{maia.vergniory@cpfs.mpg.de}
\affiliation{Donostia International Physics Center, 20018 Donostia-San Sebastián, Spain}
\affiliation{Max Planck Institute for Chemical Physics of Solids, Dresden D-01187, Germany}

\author{D.-H. Lu}
\email{dhlu@slac.stanford.edu}
\affiliation{Stanford Synchrotron Radiation Lightsource, SLAC National Accelerator Laboratory, Menlo Park, 94025, CA, USA}

\author{Z.-X. Shen} 
\email{zxshen@stanford.edu}
\affiliation{Stanford Institute for Materials and Energy Sciences, SLAC National Accelerator Laboratory, Menlo Park, 94025, CA, USA}
\affiliation{Department of Applied Physics and Physics, Stanford University, Stanford, 94305, CA, USA}
\affiliation{Geballe Laboratory for Advanced Materials, Stanford University, Stanford, 94305, CA, USA}

\date{\today}

\begin{abstract}
FeTe$_{0.55}$Se$_{0.45}$(FTS) occupies a special spot in modern condensed matter physics at the intersections of electron correlation, topology, and unconventional superconductivity. The bulk electronic structure of FTS is predicted to be topologically nontrivial due to the band inversion between the $d_{xz}$ and $p_z$ bands along $\Gamma$-$Z$. However, there remain debates in both the authenticity of the Dirac surface states (DSS) and the experimental deviations of band structure from the theoretical band inversion picture. Here we resolve these debates through a comprehensive ARPES investigation. We first observe a persistent DSS independent of $k_z$. Then, by comparing FTS with FeSe which has no band inversion along $\Gamma$-$Z$, we identify the spectral weight fingerprint of both the presence of the $p_z$ band and the inversion between the $d_{xz}$ and $p_z$ bands. Furthermore, we propose a renormalization scheme for the band structure under the framework of a tight-binding model preserving crystal symmetry. Our results highlight the significant influence of correlation on modifying the band structure and make a strong case for the existence of topological band structure in this unconventional superconductor.

\end{abstract}

\maketitle
The iron-based superconductors (IBS) have substantially contributed to our understanding of electron correlation and unconventional superconductivity\cite{si2016high, he2013phase, lee2014interfacial,qazilbash2009electronic, yi2015observation,chi2013neutron,kuo2016ubiquitous,mou2011distinct,fang2008theory,fernandes2014drives}. More recently, this material system also emerges as one of the leading platforms for the possible realization of topological superconductivity\cite{qi2011topological,zhang2019multiple,sato2017topological,hao2019topological,shi2017fete1,xu2016topological,hao2014topological, wu2016topological,liu2020new}. Among all proposed topological superconductors in IBS, FTS is the most investigated system. From density functional theory (DFT) calculations, the topological nature of FTS is predicted to arise from the inversion between the bulk $p_z$ and $d_{xz}$ bands along $\Gamma$-$Z$\cite{wang2015topological}. However, unlike weakly correlated topological materials such as Bi$_2$Te$_3$ where theoretical predictions are usually reliable\cite{hsieh2008, ylchen2009}, experimental inputs are indispensable for a comprehensive understanding of FTS because of its strong electron correlation. Scanning tunneling spectroscopy (STS) experiments pioneered in identifying the possible topological superconductivity by observing zero-energy peaks at the vortex centers which were interpreted as Majorana zero-energy modes\cite{yin2015observation, massee2015imaging, wang2018evidence, machida2019zero, chen2020atomic}. However, other STS studies found that the zero-energy peaks are unexpectedly absent in some of the vortex cores\cite{chen2018discrete,machida2019zero}. In ARPES experiments, Zhang \textit{et al.} first observed a Dirac-cone-like feature at a single photon energy and interpreted it as the DSS\cite{zhang2018observation}. Nevertheless, a later experiment reported that this feature only exists at certain photon energies, and speculated that it might originate from a bulk band instead of the DSS\cite{borisenko2020strongly}. More importantly, the measured $k_z$ dispersions\cite{johnson2015spin, thirupathaiah2016effect, lohani2020band} differ significantly from DFT predictions\cite{wang2015topological, zhang2018observation,zhang2019multiple}, with no direct evidence for either the $p_z$ orbital or the band inversion in the bulk band structure. These observations challenge the claimed topological superconductivity in FTS. To resolve these conflicts, we perform systematic ARPES measurements. 

\begin{figure*}%
\centering
\includegraphics[scale = 1]{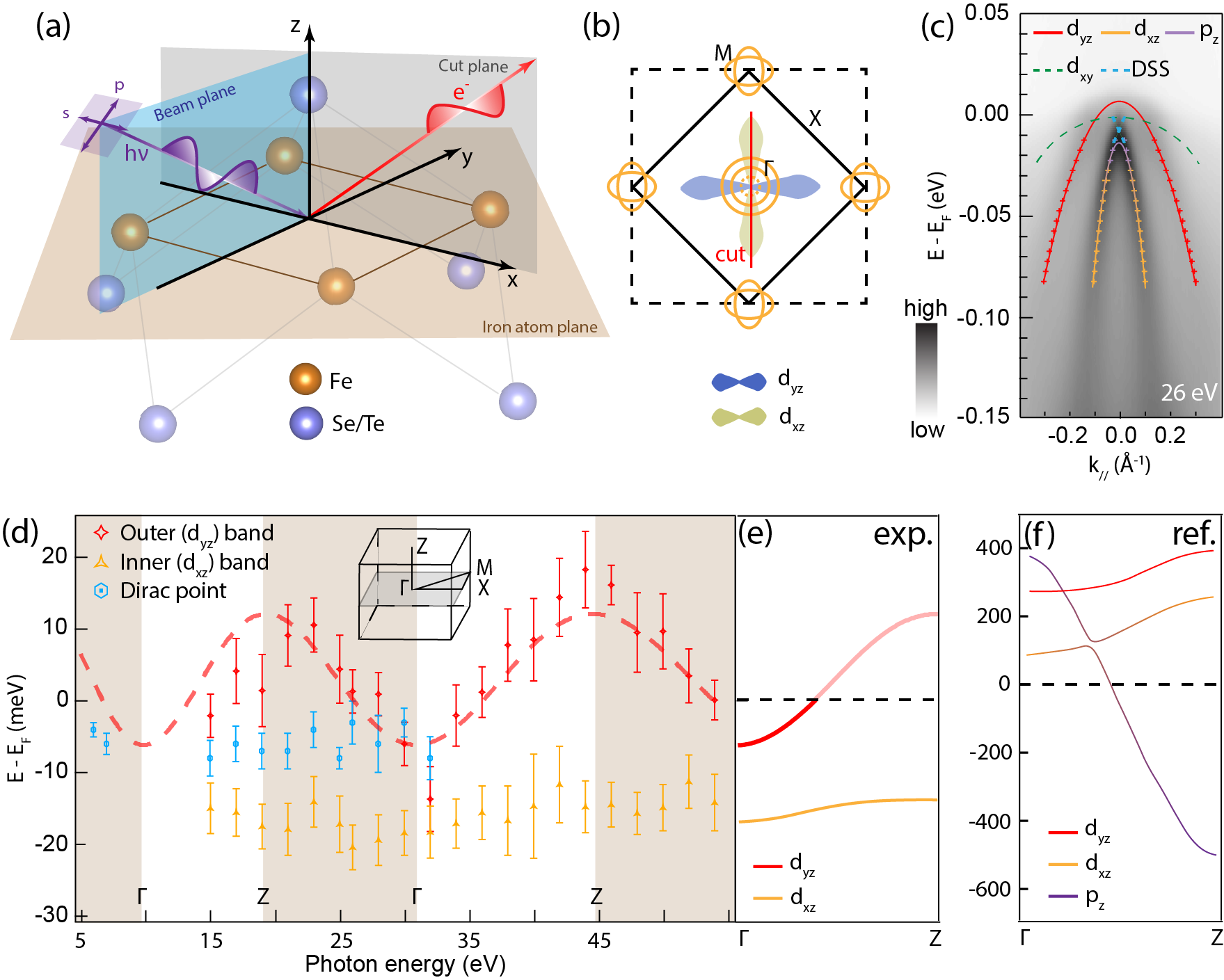}
\caption{\textbf{ARPES experimental geometry and band dispersions along $\Gamma$-$Z$. (a)} Schematic of the ARPES experimental geometry. The electric field of $s$ and $p$ polarization are normal and parallel to the beam plane, respectively. \textbf{(b)} Schematic of the cut geometry along $\Gamma$-$M$. Black solid and dashed lines show 2-Fe and 1-Fe Brillouin zones (BZ), respectively. Orange solid lines mark two hole-like pockets near the zone center ($d_{yz}$ and $d_{xy}$ bands) and two electron pockets at $M$. The orange dashed line marks one hole-like band closely below $E_F$ ($d_{xz}$ band). Green and blue patterns show the shapes of $d_{xz}$ and $d_{yz}$ orbitals projected on the $xy$-plane, respectively. \textbf{(c)} An representative ARPES spectra along $\Gamma$-$M$ at $h\nu=26$ eV with $p$ polarization. Crosses show the band dispersion extracted from momentum distribution curve (MDC) analysis. Red and purple-orange solid lines are parabolic fits to the $d_{yz}$ and $p_z$-$d_{xz}$ bands, respectively. Green and blue dashed lines are guides to the eye for the $d_{xy}$ band and Dirac surface state (DSS), respectively. \textbf{(d)} Photon energy dependence of low energy electronic states. The red dashed guide to the eye indicates the periodicity of the outer band top along $\Gamma$-$Z$. Inset shows the 3D-BZ and high symmetry points. \textbf{(e-f)} Schematic of the bulk band dispersion along $\Gamma$-Z. The dispersion extracted from (d) and adapted from ref. \cite{wang2015topological} are shown in (e) and (f), respectively.}
\end{figure*}

We start by assigning the observed features near the Brillouin zone (BZ) center in FTS to the corresponding electronic states. According to band structure calculations\cite{wang2015topological}, the states close to Fermi level are Fe $d_{xz}$, $d_{yz}$, $d_{xy}$, and Se/Te $p_z$. The intensities of the three Fe $t_{2g}$ orbitals are strongly affected by matrix element effects near the BZ center\cite{yi2015observation, sobota2021angle, zhang2011orbitalcharacter, yiming2011pnas, zkliu2015, Wang2012fematrixelemnt}. Along the $\Gamma$-$M$ direction under $p$ ($s$) polarization, only the $d_{yz}$ ($d_{xz}$) orbital is enhanced while the other $t_{2g}$ orbitals are suppressed (see supplementary information (SI) sections 1 and 6 for details). Fig. 1(c) shows a representative ARPES spectrum along $\Gamma$-$M$ under $p$ polarization at photon energy $h\nu=26$ eV (the corresponding data under $s$ polarization are shown in Fig. S1(b)). Similar to the other IBS in the 11 family\cite{zhang2011orbitalcharacter, yiming2011pnas, yi2015observation, zkliu2015}, we observe three hole bands near $E_F$. The outer band (Fig. 1(c), red line) is of $d_{yz}$ character and is only observed under $p$ polarization. The $d_{xy}$ orbital produces a flat hole-like band (Fig. 1(c), green line) with weak intensity, which can be more clearly identified in the second-energy-derivative spectra (Fig. S1(c), green dashed line). The inner band is typically identified with the $d_{xz}$ orbital in 11-family compounds and has weak (strong) intensity under $p$ ($s$) polarizations. Surprisingly, the inner band here (Fig. 1(c), orange-purple line) has strong intensity under both $p$ and $s$ polarizations, which we will explain later.

We perform parabolic fits to the in-plane dispersions of bulk bands obtained from momentum distribution curve (MDC) analysis (see SI section 2 for details). The extracted band top positions are plotted in Fig. 1(d) as a function of photon energy (or $k_z$, with fitted inner potential $V_0\approx 11.3$ eV, see SI section 3 for details). The outer band top shows periodic oscillations spanning a $\sim$20 meV range (red guideline in Fig. 1(d)), while the inner band shows almost no $k_z$ dispersion within measurement uncertainties, consistent with previous reports\cite{johnson2015spin, thirupathaiah2016effect, lohani2020band}. Fig. 1(e) and (f) shows the schematic band $k_z$ dispersion extracted from experiments and the DFT calculations in ref. \cite{wang2015topological}, respectively. Contrary to the DFT calculations, first glance revealed neither a $k_z$-dispersive $p_z$ band nor band inversion, with an additional inconsistency in Fermiology. Indeed, the $k_z$ dispersion of the inner band is almost identical to those in other 11-family compounds without band inversion\cite{subedi2008density}, raising questions on the predicted topological nature in FTS.

To resolve the discrepancies, in the following, we will report three observations unique to FTS to support a topological band structure: (i) the $k_z$-independent DSS, (ii) the presence of $p_z$ orbital character in the inner band, and (iii) the orbital character cross-over hinted by the spectral weight variations.

\begin{figure*}%
\centering
\includegraphics[scale = 1]{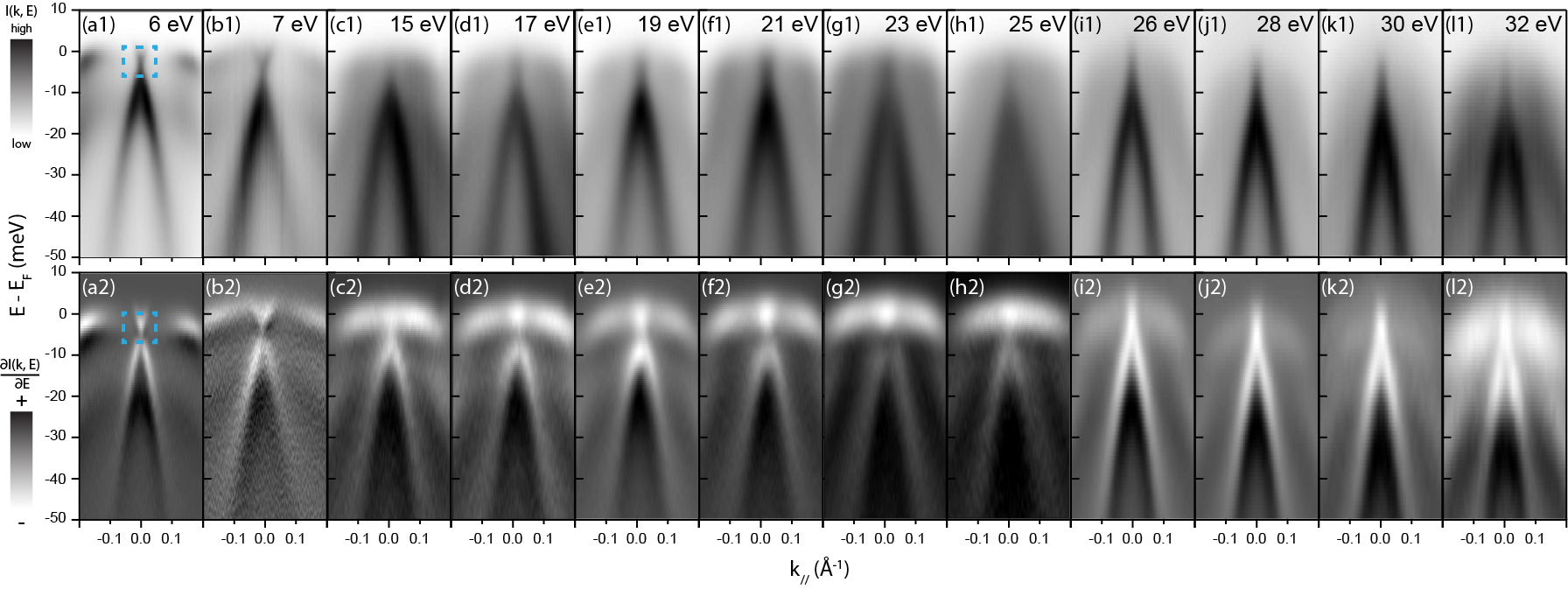}
\caption{\textbf{Dirac surface state (DSS) at different photon energies. (a1-l1)} ARPES spectra over a wide range of photon energies. \textbf{(a2-l2)} The first-energy-derivative spectra in a1-l1. Blue dashed boxes indicate the energy-momentum window containing the DSS signal. 6-7 eV spectra are taken with lasers. 15-25 eV and 26-32 eV spectra are taken with synchrotron light source at two complementary beamlines. All cuts go through BZ center.}
\end{figure*}

We first demonstrate the $k_z$-independent DSS. Fig. 2(a1)-(l1) show the zoom-in spectra near $E_F$ over a wide range of photon energies, taken with 6 eV and 7 eV lasers and synchrotron photons between 15 eV and 32 eV. Our result is consistent with the first report of the DSS at $h\nu=7$ eV\cite{zhang2018observation}, but contains data covering the entire $\Gamma$-$Z$ trajectory. Contrary to the previous photon-energy-dependence study\cite{borisenko2020strongly}, we find that the DSS feature persists throughout the photon energy range in Fig. 2(a1)-(l1) in the energy-momentum window indicated by the blue dashed box (Fig. 2(a1, a2)). The first-energy-derivative plots in Fig. 2(a2)-(l2) better highlight the DSS at some photon energies. We note that the DSS feature exists in a very narrow momentum and energy window, and consequently requires careful measurements with optimized measurement alignment, small beam spot size, and high instrument resolution. Following an earlier study\cite{lohani2020band}, we further perform a quantitative analysis to confirm the $k_z$-independent DSS signal. We fit MDCs between -14 meV and $E_F$, and extract the widths of the MDC peaks associated with the DSS as a function of energy (see SI section 4 for details). The MDC widths are expected to reach minimum at the binding energy of the Dirac point. In Fig. 1(d), we show that the binding energy of the Dirac point stays fixed as a function of photon energy, despite the $\sim$ 20 meV $k_z$ dispersion of the $d_{yz}$ band. This confirms the 2D nature of the DSS. Above 32 eV, the DSS signal becomes hard to identify because of the deteriorated in-plane momentum resolution and a potential matrix element suppression of the $p_z$ atomic orbital component (to be discussed below)\cite{peng2019FTSfilms}. Nevertheless, the photon energy range presented here is large enough to cover the entire $\Gamma$-$Z$ range (Fig. 1(d)). Therefore, our data supports the DSS interpretation and exclude the possibility that the Dirac-cone-like dispersion comes from the subtle bulk-band crossing at specific $k_z$\cite{borisenko2020strongly}.

Having confirmed the existence of DSS, we now look for its bulk correspondence by first searching for the evidence of the $p_z$ orbital that participates in the band inversion with the $d_{xz}$. In Fig. 3, we compare the spectra of FTS with those of FeSe, a closely related compound predicted to have no band inversion along $\Gamma$-$Z$\cite{wang2015topological}. Fig. 3(a-e) and Fig. 3(g-i) show spectra of FTS and FeSe along the $\Gamma$-$M$ direction at different photon energies, respectively. We first compare the band intensity as a function of in-plane momentum ($k_\parallel$). Along the $\Gamma$-$M$ direction under $p$ polarization, the matrix element of the $d_{xz}$ is suppressed near $k_\parallel=0$, while the matrix elements of $d_{yz}$ and $p_z$ orbitals remain large (Fig. S10, S12). Experimentally, we indeed find that the inner band of FeSe, whose orbital character is dominantly $d_{xz}$ without considering the SOC\cite{subedi2008density}, has suppressed intensity near the inner band top ($k_\parallel=0$) (Fig. 3(g-i)). In contrast, the inner band of FTS shows maximal intensity at $k_\parallel=0$ (Fig. 3(a-e)). This indicates the presence of either the $d_{yz}$ or the $p_z$ orbital component near the inner band top.

To better understand this orbital component, we further look into the spectral weight contrast between the inner and outer bands. We quantify the contrast by defining the spectral weight ratio (SWR) as the ratio between the average intensities in region 2 (orange bar in Fig. 3(a) sampling the inner band) and region 1 (red bars in Fig. 3(a) sampling the outer band, see SI section 5 for details). The SWRs for the $\Gamma$-$M$ cut are plotted in Fig. 3(j) as a function of photon energy (see Fig. S6 and S8 for the raw spectra). For FeSe, the SWR remains almost constant with tiny modulations. This behavior is expected because the dominant orbital components for the inner and outer bands in FeSe are $d_{xz}$ and $d_{yz}$, respectively\cite{subedi2008density}, and the matrix element ratio between these two orbital components is roughly photon-energy independent (Fig. S10). In contrast, the SWR for FTS shows strong and intricate modulations, echoing the in-plane intensity distribution anomaly and consistent with the $k_z$-dependent mixing of either the $d_{yz}$ or the $p_z$ orbital components in the inner $d_{xz}$ band. In the case of $d_{yz}$ mixing with $d_{xz}$, the SWR modulation is only expected in the data collected along the $\Gamma$-$M$ but not the $\Gamma$-$X$ direction, because these two orbitals have the similar matrix elements along $\Gamma$-$X$ as required by symmetry (Fig. S5(d), also see SI section 5 for details). However, we also observe a similar SWR modulation in data taken along $\Gamma$-$X$ (see SI section 5 for details). Therefore, the SWR modulations can only arise from the hybridization between the $d_{xz}$ and $p_z$ orbitals near the inner band top. The contrasting behaviors in FeSe and FTS highlight the importance of the Te $5p_z$ orbital: it introduces strong inter-layer hopping and brings the $p_z$ dispersion across the Fe $t_{2g}$ bands\cite{wang2015topological}, eventually leading to a significant mixing of the $p_z$ orbital character into the inner Fe $d_{xz}$ band.

\begin{figure*}%
\centering
\includegraphics[scale = 1]{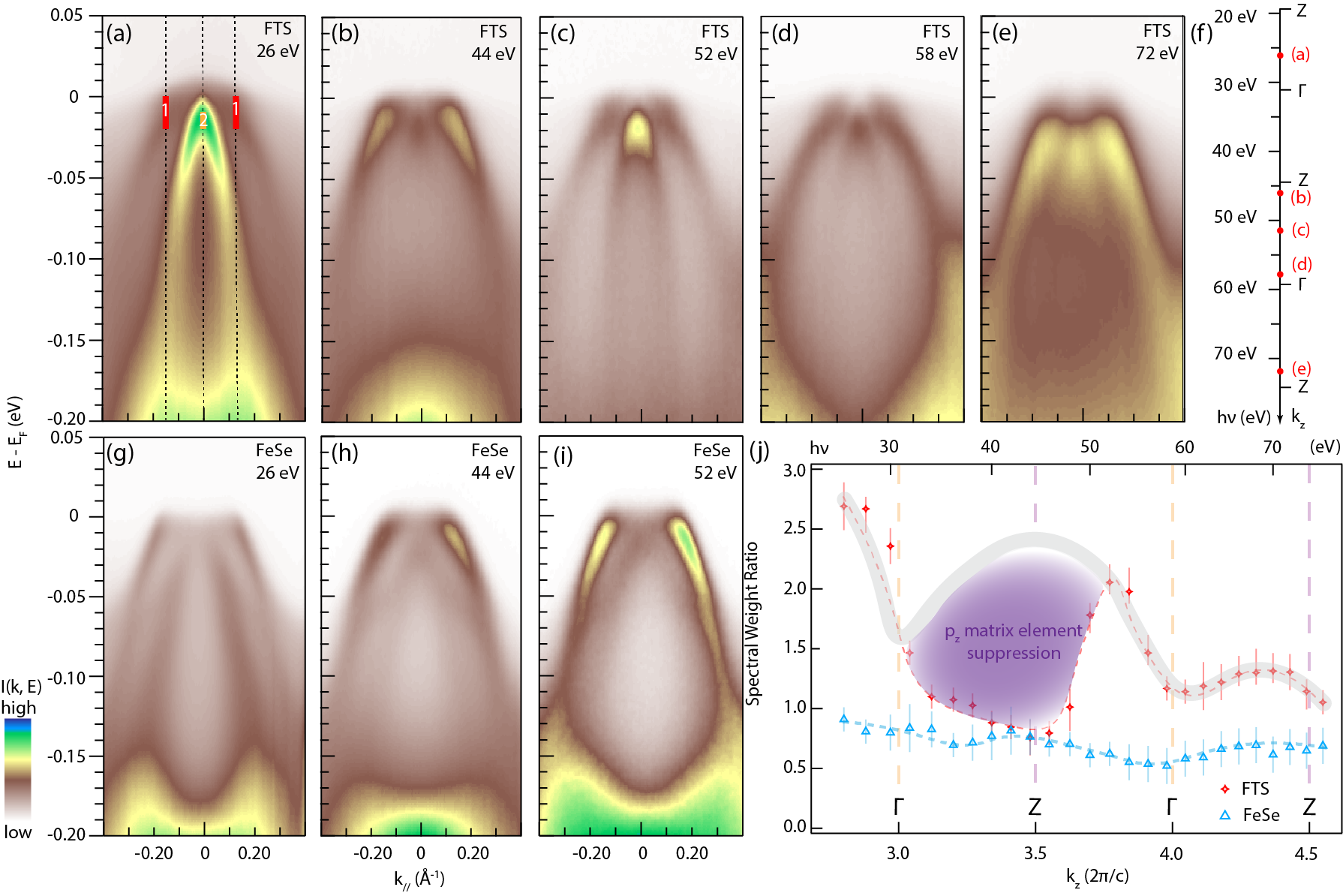}
\caption{\textbf{Spectral weight modulation in FTS. (a-e)} FTS spectra taken at $h\nu=$ (a) 26 eV, (b) 44 eV, (c) 52 eV, (d) 58 eV, (e) 70 eV, respectively. The red and orange bars in (a) indicate region 1 and region 2 in spectral weight analysis, respectively. \textbf{(f)} Correspondence between photon energy and $k_z$ for FTS. Red dots mark photon energies in (a-e). \textbf{(g-i)} FeSe spectra taken at (g) 26 eV, (h) 44 eV, (i) 52 eV, respectively. \textbf{(j)} The spectral weight ratio (SWR) as a function of photon energy for FTS and FeSe. Orange and purple dashed lines indicate $\Gamma$ and $Z$ points, respectively. The grey line is a guide to the eye for the expected oscillatory behavior. The purple shaded area marks the anomalous matrix element depression. All cuts are along $\Gamma$-$M$ taken under $p$ polarization.}
\end{figure*}

Having established the $p_z$ orbital component, we now further look for evidence of $d_{xz}$-$p_z$ orbital character crossover near the inner band top as a function of $k_z$ -- a natural consequence of the bulk band inversion. The SWR in FTS in Fig. 3(j) can be understood as a product of two effects: an oscillatory behavior by the orbital character change (grey guide to the eye) and an anomalous depression of the photoemission matrix element near 40 eV (purple shaded area). The overall oscillatory behavior suggests that the spectral weight of the inner band reaches the maximum at $Z$ and the minimum at $\Gamma$. This implies a $d_{xz}$-$p_z$ crossover of the orbital character going from $\Gamma$ to $Z$ at the inner band top since the $p_z$ orbital has a much larger photoemission matrix element than that of the $d_{xz}$ orbital under $p$ polarization. On the other hand, the anomalous suppression can be attributed to the matrix element effect. To show this, we calculate the photoemission dipole matrix element approximately using free-electron plane-wave states as final states and hydrogen-like wave functions as initial states (see SI section 6.1 for details)\cite{moser2017experimentalist, wang2012orbital, day2019computational}. The calculated matrix elements of the $d_{xz}$ and $p_z$ orbitals are further normalized by that of the $d_{yz}$ orbitals for direct comparison with SWR (Fig. S11). We observe a strong atomic photoemission matrix element suppression of the $p_z$ orbital around 40 eV, consistent with the experimental data. However, we cautiously note that such matrix element suppression is sensitive to the approximations of wave functions for the initial and final states (see SI section 6 for detailed discussion). Nevertheless, the novel spectral weight in FTS, highlights the existence of $p_z$ orbital, and furthermore, serves as a strong hint for the $d_{xz}$-$p_z$ orbital character crossover.

We remark that a related spectral weight modulation of the $d_{xz}$ orbital in the inner band under the $s$ polarization has been reported in Ref. \cite{lohani2020band} with a similar $k_z$ periodicity in support of band inversion. However, our comprehensive data here reveal the more critical and unconventional elements in FeSCs, the $p_z$ orbital component under the $p$ polarization, and present a complete picture with more convincing experimental evidence indicating the inversion between the $p_z$ and $d_{xz}$ bands by (i) the identification of the $p_z$ orbital and exclusion of other mixing possibilities in FTS, (ii) the expected absense of such $p_z$ orbital in the control experiment of FeSe, (iii) the observation of SWR modulation further supporting the $d_{xz}$ and $p_z$ band crossing along the $\Gamma$-$Z$ direction, and (iv) the persistent DSS required by a non-trivial topological band structure.

With a confirmed topological band structure, we finally show a picture to reconcile the discrepancy between Fig. 1(e) and (f) under the framework of a tight-binding model preserving crystal symmetry (see SI section 7 for details). Starting from the band structure calculated by the DFT (Fig. 4(a)), we make adjustments in two steps: (i) renormalize the $d$ bands by a factor of 3 with respect to the experimental $E_F$ in accordance with the in-plane dispersion\cite{qazilbash2009electronic, yi2015observation, ymingreview2017, zkliu2015} (Fig. 4(b)); (ii) increase the $d_{xz}$-$p_z$ SOC strength as a phenomenological factor to further capture the many-body interactions (Fig. 4(c)). See SI section 7 for details. It is worth noting that a quantitative reproduction of experimental observations further requires a $E_F$ that locates in the gap between the $d_{xz}$ and $p_z$ bands (Fig. 4(d)), and a reduction of energy separation between the $d_{yz}$ and $d_{xz}$ bands. We note that the relative position of the hole-like band with respect to the Fermi-level also varies in literature\cite{wang2015topological, zhang2018observation}. The origin of this discrepancy remains an ongoing research topic, with strong correlations and Mott transition effects as potential candidates\cite{xiaoboma2022, minjaekim2023,zhiguangliao2023}. Nevertheless, our picture vividly highlights the critical role of strong correlation in the realization of an overall topologically non-trivial band character by dramatically renormalizing the DFT predictions. This greatly reduces the energy scale and ensures the proximity of the inverted band gap to the $E_F$ for topological physics to be relevant.

\begin{figure}[b]
\includegraphics[scale=1.20]{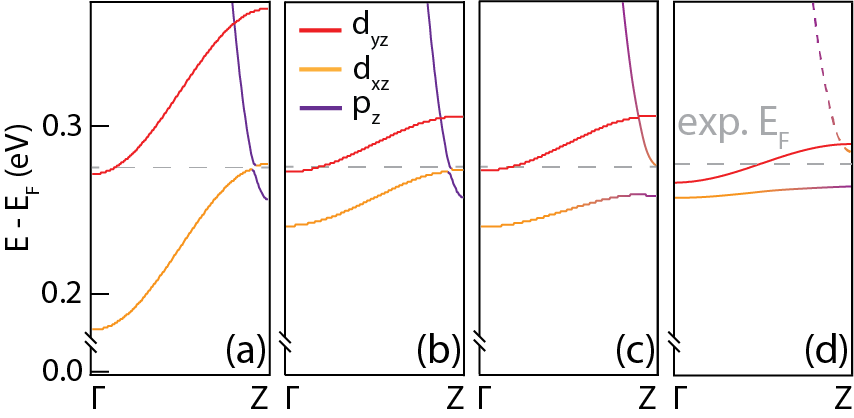}
\caption{\textbf{Renormalization steps for the band structure.} (a) DFT calculation results. (b) Renormalize the $d$ bands by a factor of 3 with respect to the experimental $E_F$ (grey dashed line). (c) A further increase of $d_{xz}$-$p_z$ SOC strength as a phenomenological measure of many-body correlations. (d) Comparison with the experimental data with parameters extracted from Fig. 1(d).}
\end{figure}

In conclusion, we provide a strong case for a topological band structure in FeTe$_{0.55}$Se$_{0.45}$ by systematically analyzing the band dispersion and spectral weight. Our picture reconciles the discrepancy between the calculated band structure and experimental observations by necessary many-body renormalizations and provides a baseline for the further pursuit of topological physics in this system.

\textbf{Experimental methods.} FTS samples were grown by flux method\cite{liu2009samplegrowth}. Magnetic susceptibility measurements show superconducting transition at 14 K, in agreement with the previous report\cite{zhang2018observation}. The FTS data at 6 eV and 7 eV photon energies were collected using laser APRES systems at Stanford University, with Scienta R4000 and R8000 analyzers, respectively. The energy resolution was 5 meV (3 meV) and the measurement temperature was 20 K (16 K) for the 6 eV (7 eV) experiments. FTS data between 15 eV and 25 eV data were collected at the Stanford synchrotron radiation lightsource (SSRL) beamline 5-4 using a Scienta R4000 analyzer, with energy resolution at 5 meV and temperature at 16 K. The FTS data between 26 eV and 76 eV were collected at SSRL beamline 5-2 using a Scienta DA 30 analyzer with energy resolution at 12 meV and temperature at 16 K. The FeSe data between 26 eV and 52 eV are collected at SSRL beamline 5-2, on a detwinned sample\cite{pfau2019momentum}, with energy resolution at 12 meV and temperature at 12 K. The pressure was kept below $4\times 10^{-11}$ Torr throughout all ARPES measurements. 

\textbf{Supplemental information.} Photoemission selection rule and matrix element calculations are discussed in detail in SI sections 1 and 6, respectively. Bulk band $k_z$ dispersion fit, conversion between photon energies and $k_z$, analysis of Dirac surface state signal, spectral weight analysis are discussed in details in SI section 2, 3, 4, and 5, respectively. Details for the DFT calculations and the tight-binding model are shown in SI section 7.

\textbf{Acknowledgement.} We acknowledge Y. He, T.P. Devereaux, D.-H. Lee, S. Kivelson, B. Moritz, and X. Dai for useful discussions. Synchrotron ARPES measurements were performed at Beamline 5-2 and 5-4, Stanford Synchrotron Radiation Lightsource, SLAC National Accelerator Laboratory. The works at Stanford University and SLAC are supported by the U.S. Department of Energy, Office of Science, Office of Basic Energy Sciences, Division of Materials Sciences and Engineering, under Contract No. DE-AC02-76SF00515. Support for FTS crystal growth and characterization at Penn State is provided by the National Science Foundation through the Penn State 2D Crystal Consortium-Materials Innovation Platform (2DCC-MIP) under NSF Cooperative Agreement DMR-2039351. The FeSe single crystal growth at Rice is supported by the US DOE, BES under grant No. DE-SC0012311. M.G.V. thanks support to the Deutsche Forschungsgemeinschaft (DFG, German Research Foundation) GA 3314/1-1 – FOR 5249 (QUAST) and partial
support from European Research Council (ERC) grant agreement no. 101020833.

\textbf{Author contributions.} Y.F.L., D.H.L., and Z.X.S. designed the experimental plan. Y.F.L., S.D.C., H.P., and J.A.S. performed ARPES experiments and data analysis. Y.F.L. and H.P. performed matrix element calculations. M.G.D., M.I.I., and Y.F.L. performed DFT calculations and tight-binding modeling. Y.L.Z. and Z.Q.M. synthesized high-quality FTS samples. T.C., M.Y., and P.C.D. synthesized high-quality FeSe samples. Y.F.L., S.D.C., M.G.D., M.G.V., D.H.L., and Z.X.S. wrote the manuscript with inputs from all authors. J.A.S., M.H., D.H.L., and Z.X.S. supervised the whole project.


\end{document}


\preprint{APS/123-QED}
\title{Supplementary information: Spectroscopic evidence for topological band structure in FeTe$_{0.55}$Se$_{0.45}$}

\author{Y.-F. Li}
\thanks{These authors contributed equally.}
 \affiliation{Stanford Institute for Materials and Energy Sciences, SLAC National Accelerator Laboratory, Menlo Park, 94025, CA, USA}
\affiliation{Department of Applied Physics and Physics, Stanford University, Stanford, 94305, CA, USA}
\affiliation{Geballe Laboratory for Advanced Materials, Stanford University, Stanford, 94305, CA, USA}

\author{S.-D. Chen}
\thanks{These authors contributed equally.}
\affiliation{Stanford Institute for Materials and Energy Sciences, SLAC National Accelerator Laboratory, Menlo Park, 94025, CA, USA}
\affiliation{Department of Applied Physics and Physics, Stanford University, Stanford, 94305, CA, USA}
\affiliation{Geballe Laboratory for Advanced Materials, Stanford University, Stanford, 94305, CA, USA}
\affiliation{Department of Physics, University of California, Berkeley, California 94720, USA}

\author{M. Garcia-Diez}
\affiliation{Donostia International Physics Center, 20018 Donostia-San Sebastián, Spain}
\affiliation{Physics Department, University of the Basque Country (UPV/EHU), Bilbao, Spain}

\author{M. I. Iraola}
\affiliation{Donostia International Physics Center, 20018 Donostia-San Sebastián, Spain}

\author{H. Pfau}
 \affiliation{Stanford Institute for Materials and Energy Sciences, SLAC National Accelerator Laboratory, Menlo Park, 94025, CA, USA}
\affiliation{Advanced Light Source, Lawrence Berkeley National Laboratory, Berkeley, 94720, CA, USA}
\affiliation{Department of Physics, Pennsylvania State University, University Park, 16802, PA, USA}

\author{Y.-L. Zhu}\author{Z.-Q. Mao}
\affiliation{Department of Physics, Pennsylvania State University, University Park, 16802, PA, USA}

\author{T. Chen}\author{M. Yi}\author{P.-C. Dai}
\affiliation{Department of Physics and Astronomy, Rice University, Houston, 77005, TX, USA}

\author{J.A. Sobota}
 \affiliation{Stanford Institute for Materials and Energy Sciences, SLAC National Accelerator Laboratory, Menlo Park, 94025, CA, USA}

\author{M. Hashimoto}
\affiliation{Stanford Synchrotron Radiation Lightsource, SLAC National Accelerator Laboratory, Menlo Park, 94025, CA, USA}

\author{M. G. Vergniory}
\email{maia.vergniory@cpfs.mpg.de}
\affiliation{Donostia International Physics Center, 20018 Donostia-San Sebastián, Spain}
\affiliation{Max Planck Institute for Chemical Physics of Solids, Dresden D-01187, Germany}

\author{D.-H. Lu}
\email{dhlu@slac.stanford.edu}
\affiliation{Stanford Synchrotron Radiation Lightsource, SLAC National Accelerator Laboratory, Menlo Park, 94025, CA, USA}

\author{Z.-X. Shen} 
\email{zxshen@stanford.edu}
\affiliation{Stanford Institute for Materials and Energy Sciences, SLAC National Accelerator Laboratory, Menlo Park, 94025, CA, USA}
\affiliation{Department of Applied Physics and Physics, Stanford University, Stanford, 94305, CA, USA}
\affiliation{Geballe Laboratory for Advanced Materials, Stanford University, Stanford, 94305, CA, USA}

\date{\today}

\maketitle
\tableofcontents
\section{Photoemission selection rule}
Under the dipole approximation, the photoemission matrix element is given by
\begin{equation}
  M_{f,i} \propto \bra{f}\hat{\boldsymbol\epsilon} \cdot \boldsymbol r\ket{i}
\end{equation}
, where $\ket{f}$ is the final state, $\hat{\boldsymbol\epsilon}$ is a unit vector representing the electric field of the light, $\boldsymbol r$ is the spatial vector, and $\ket{i}$ is the initial state.

Here we use hydrogen-like wavefunctions and plane-wave free-electron states to approximate the initial and final states, respectively. For cut along $\Gamma$-$M$ near the zone center, a set of qualitative selection rules can be obtained by considering the symmetry of the integrand in Eq. (1) with respect to either the cut plane and the beam plane (Fig. 1(a)). We summarize the results in Table S1, where E/S stands for enhanced (suppressed), NA stands for not applicable and $+(-)$ represents the even (odd) parity with respect to the reference plane.

For cuts along $\Gamma$-$X$ at zone center (Fig. S5(d)), both $d_{xz}$ and $d_{yz}$ orbitals are ${45}^\circ$ offset from the cut plane and beam plane. Therefore, these two orbitals have similar photoemission intensities along $\Gamma$-$X$. The conclusions here are also supported by the calculation in SI section 6.

\begin{table*}[h]
\begin{center}
\caption{TABLE S1. Parity analysis of photoemission matrix element}
\begin{tabular}{p{2cm}cc|p{2cm}cc}
\toprule
 &\multicolumn{2}{p{4cm}|}{\textbf{Cut plane}} & &\multicolumn{2}{p{4cm}}{\textbf{Beam plane}} \\
\hline
 & $p$ (mixed)  & $s$ (+) & & $p$ (+)  & $s$ (-)\\
\hline
$d_{xz}$(+) & NA & E  & $d_{xz}$(-) & S & E\\
$d_{yz}$(-) & NA & S & $d_{yz}$(+) & E & S \\
\botrule
\end{tabular}
\end{center}
\end{table*}

\section{Bulk band dispersion}
Fig. S1(d) shows the in-plane momentum distribution curves (MDCs) from Fig. S1(a) at different binding energies. A representative MDC fit is shown in Fig. S1(e) where four Lorentz peaks and a parabolic background are used. The peak positions are marked as crosses in Fig. 1(c). We further perform a parabolic fit to the extracted peaks positions from the outer and inner bands. The band top energies obtained from the parabolic fits are plotted in Fig. 1(d). We carry out similar analysis on four independent data sets from different samples with the same nominal doping. The error bars in Fig. 1(d) indicate the range of results from the four different data sets.
\begin{figure*}[h]
\centering
\includegraphics[]{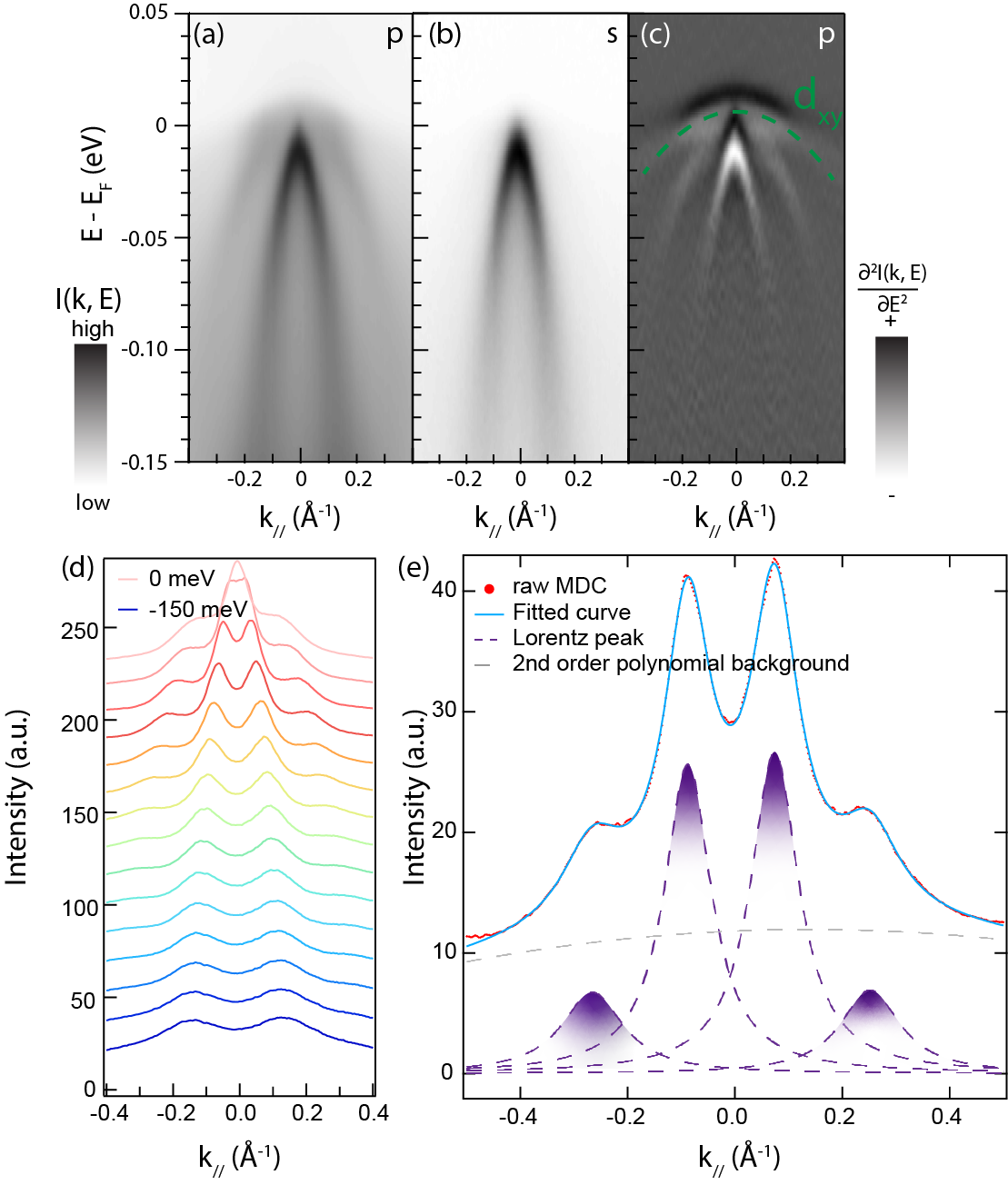}
\caption{\textbf{Fig. S1: FTS spectra along $\Gamma$-$M$ direction taken at $h\nu = 26$ eV. (a), (b)} Spectra taken under the $p$ and $s$ polarizations, respectively. The outer band appears under $p$ polarization while the inner band appears under both $p$ and $s$ polarization. \textbf{(c)} The second-energy-derivative spectra from (a). The green guide to the eye marks the $d_{xy}$ band. \textbf{(d)} The momentum distribution curves (MDCs) from (a) at binding energies from -150 to 0 meV. \textbf{(e)} A representative MDC fit at $E-E_F =$ -55 meV.}\label{fig1}
\twocolumngrid
\end{figure*}

\section{Conversion between photon energy and $k_z$}
For normal emission with $k_\parallel=0$, the map between photon energy and $k_z$ is given by
\begin{equation}
\hbar k_z = \sqrt{2m(h\nu - \phi - E_B + V_0)}
\end{equation}
, where $m$ is the electron mass, $V_0$ is the inner potential, $\phi$ is the work function, and $E_B$ is the binding energy of the electron prior to emission. We also assume the following dispersion for the outer band top
\begin{equation}
    E_B = A\cos(c\cdot k_z) + B
\end{equation}
as a function of $k_z$, where $A, B$ are the fitting parameters, and $c=5.983 \text{\AA}$ for FTS.\\\\
Combining the eq. (2) and (3), we have 
\begin{equation}
    E_B(h\nu; A, B, V_0) = A\cos(\frac{c}{\hbar}\sqrt{2m(h\nu - \phi + V_0)}) + B
\end{equation}
, where $\Phi$ is 4.35 eV in our experiment, and $E_B$ is dropped on the right hand side because $E_B\ll h\nu - \Phi + V_0$. We then perform a least-squares fit to the data (outer band top) in Fig. 1(d) using eq. (4). The red dashed line in Fig. 1(d) shows the fitting result with $A = -9.1$ meV, $B = 2.9$ meV, and $V_0 = 11.3$ eV. We note that the fitting model implicitly assumes a free-electron final state\cite{sabota2021}, which may not be accurate enough for low photon energies. This may explain the slight difference between data and fitting in the low photon energy region (15-20 eV) in Fig. 1(d).

\section{Analysis of Dirac surface state signal}
We present more analysis to demonstrate the existence of DSS signal over a wide range of photon energies (15-32 eV). Our analysis is based on two criteria: 

(1) \textbf{The DSS-related peaks have spectral weight extending towards $E_F$ in in-plane MDCs.} In Fig. S2, we show MDCs near $E_F$ at different photon energies. In addition to the two peaks from the bulk outer band, we observe a peak at $k_\parallel=0$ consistently extending towards $E_F$ at each photon energy.

\textbf{(2) The width of the DSS-related peak reaches minimum at the binding energy of the Dirac point.} In Fig. S3(a), we show a representative fit to a MDC curve with a linear background and three Lorentz peaks. The two outer peaks and the center peak originate from the bulk band and the DSS, respectively. In Fig. S3(b-k), we show the full width at half maximum of the center peak as a function of binding energy at each photon energy. The uncertainty of the width-minimum energy in Fig. 1(d) are dominated by the uncertainty of $E_F$.

We note that while the first criterion is straightforward, it might be affected by the intensity leakage from the bulk inner band. However, a mere intensity leakage without the DSS would fail the second criterion. Thus, we combine the two criteria to showcase the persistence of the DSS.

\begin{figure*}[h]
\centering
\includegraphics[scale=0.8]{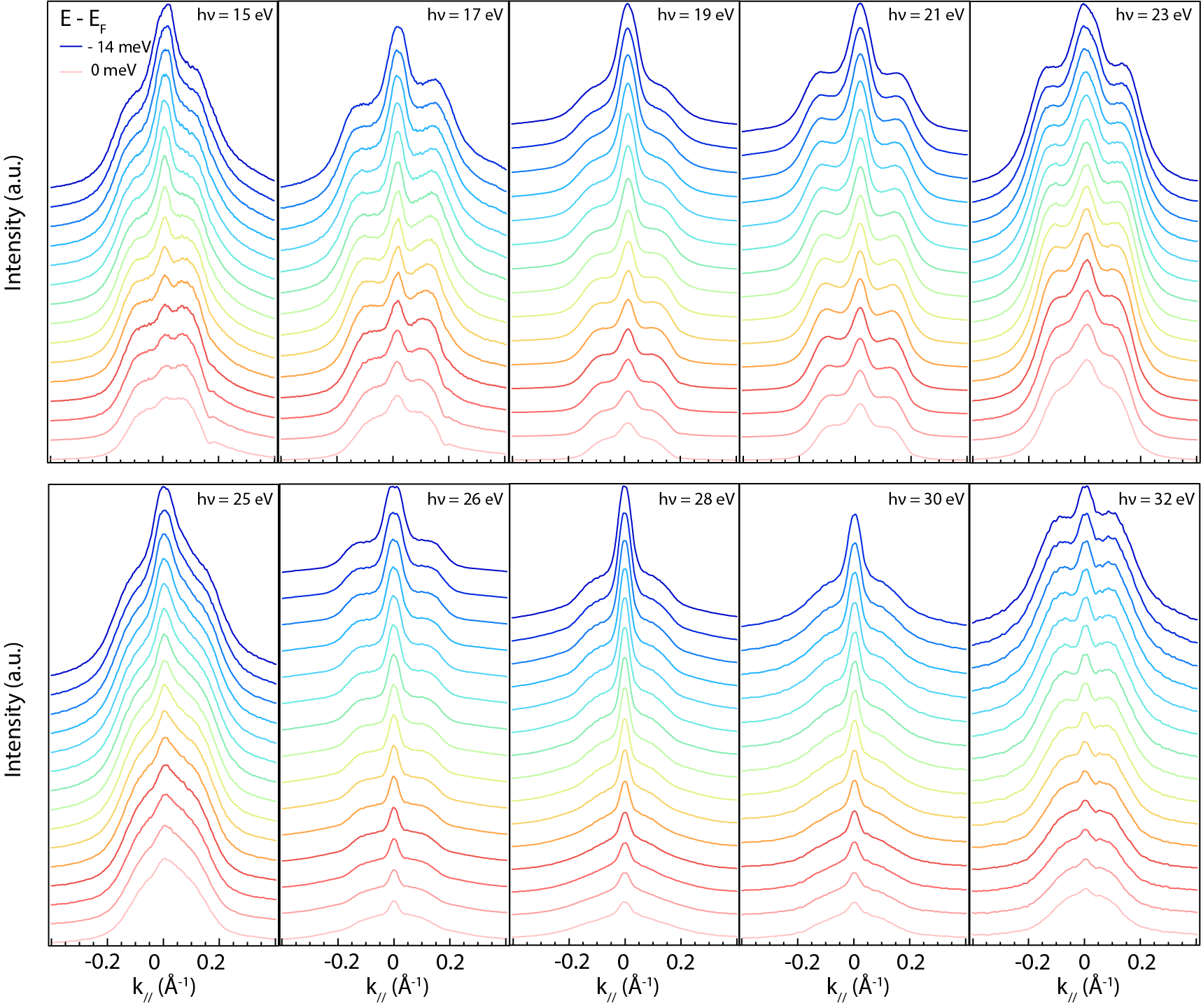}
\caption{\textbf{Fig. S2: MDCs near $E_F$ for the Dirac surface state (DSS) signal at different photon energies.} The curves are obtained from $-14$ meV (top) to $0$ meV (bottom) with even energy steps.}
\end{figure*}

\begin{figure*}[h]
\centering
\includegraphics[scale=0.8]{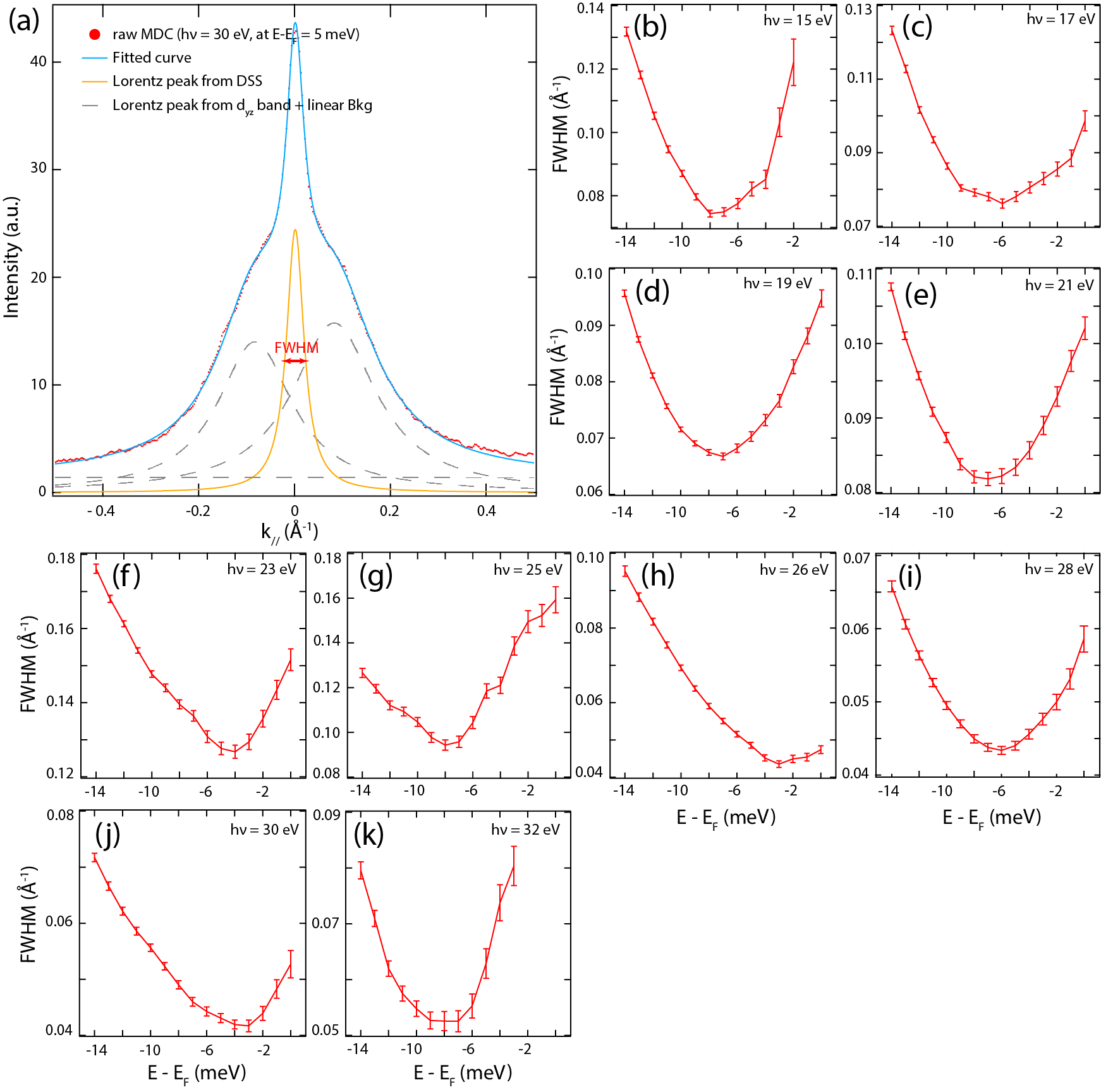}
\caption{\textbf{Fig. S3: Width analysis of DSS peak. (a)} Representative fit of MDC curve at $E-E_F=5$ meV, $h\nu=30$ eV. The red arrow marks the full width at half maximum (FWHM) of the DSS peak. \textbf{(b)-(k)} FWHM of DSS-related peak as a function of binding energy at different photon energies.}
\end{figure*}

\section{Spectral weight analysis}
Here we explain in detail the spectral weight analysis used in Fig. 3(j). Our goal is to track as a function of photon energy the spectral intensity ratio between the inner band and the outer band. Because the band dispersion is $k_z$-dependent, this ratio cannot be evaluated using sampling windows fixed in energy and $k_\parallel$. Instead, we use floating windows to track the intensities of two well-defined features, the inner band top and the outer band $k_F$. The spectral weight ratio (SWR) plotted in the main text is defined as the ratio between them.

In Fig. S4 (a1) and (b1) we show two representative ARPES images along the $\Gamma$-$M$ direction in FTS, taken using 26 and 44 eV photons, respectively. To quantify the intensity near the outer band $k_F$, we plot the $k_F$ EDCs and find their average intensities in the fixed energy window shown in Fig. S4 (a2) and (b2). The intensity is slightly asymmetric with respect to $k_\parallel=0$ likely because of the small angular offset of the sample. Thus, we further average over each pair of $k_F$ EDCs on the $k_\parallel>0$ and $k_\parallel<0$ sides to improve the accuracy of our analysis. For the inner band, we first obtain the EDC at $k_\parallel=0$, then use an averaging window with fixed energy width centered at the EDC maximum to obtain the representative intensity (see for example Fig. S4 (a3) and (b3)). From these intensities, we obtain the SWR of FTS plotted in Fig. 3(j). This analysis is performed on three independent data sets from different samples with the same nominal composition. The error bars in Fig. 3(j) represents the range of results from these data sets.

A similar analysis is performed on detwinned FeSe data sets (see for example bottom panels in Fig. S4) except that the $k_\parallel=0$ EDC is averaged in a finite momentum window between the orange dashed lines in Fig. S4(c1) and (d1) to improve the signal-to-noise ratio and compensate for spectral weight suppression at $k_\parallel=0$. This analysis is performed on one data set taken on a detwinned sample\cite{pfau2019momentum}. The error bars in Fig. 3(j) represent range of results due to uncertainties in $k_F$ determination.

The FTS data along the $\Gamma$-$X$ direction are analyzed using the same method as those along the $\Gamma$-$M$ direction, except that the $k_F$ EDCs of the outer band are averaged in fixed-width momentum windows centered at $k_F$ (as indicated by the red and blue dashed lines in Fig. S5(a1) and (b1)). This change is necessary because the spectral features are broader along the $\Gamma$-$X$ direction. The detailed steps are shown in the top panels of Fig. S5, and the obtained SWR is plotted in Fig. S5 (c). The analysis is performed on one data set and the error bars represent range of results due to uncertainties in $k_F$ determination.

Representative raw data used in the analysis are shown in Fig. S6, S7, S8.
\begin{figure*}[h]%
\centering
\includegraphics[scale=0.8]{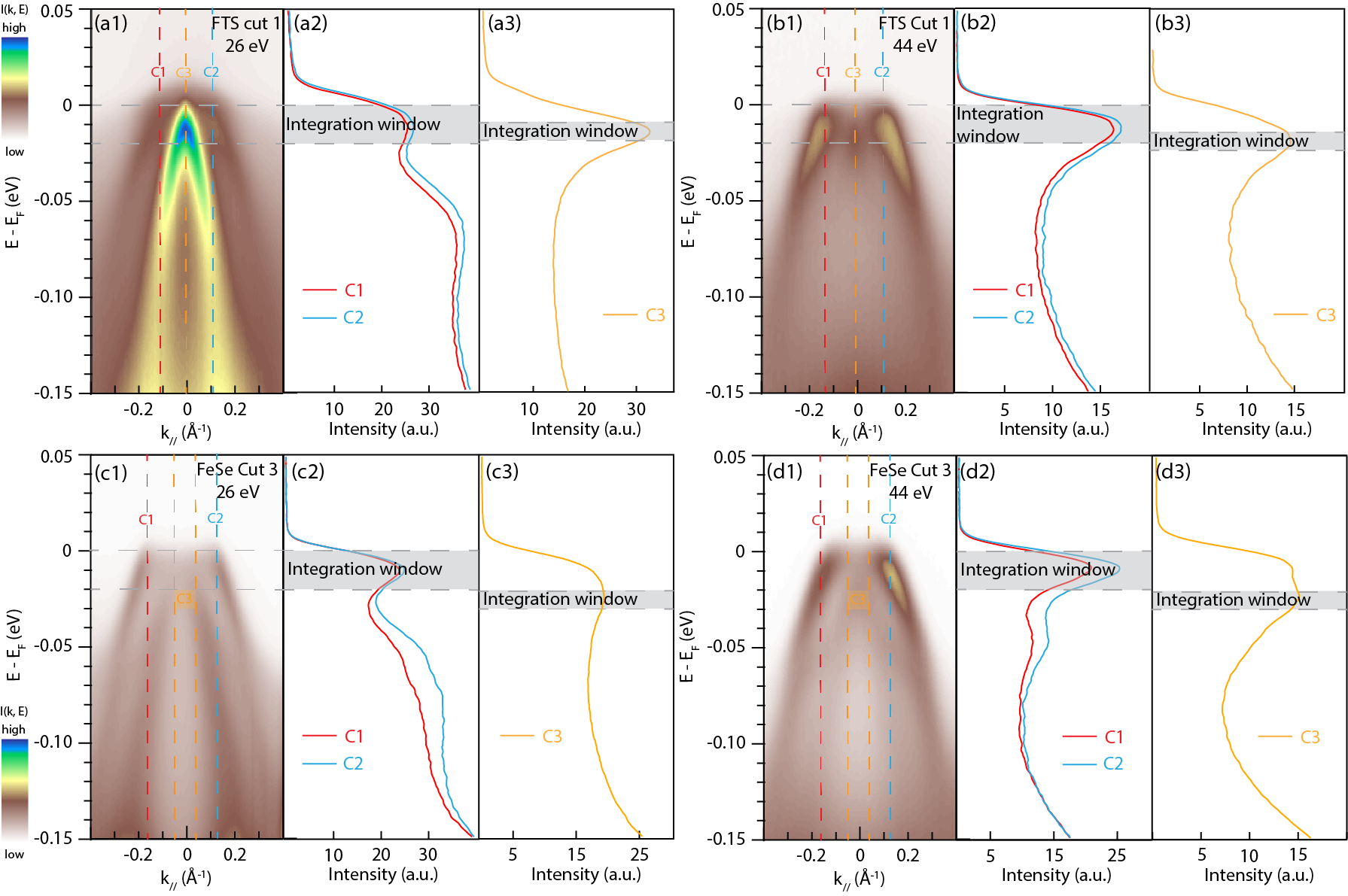}
\caption{\textbf{Fig. S4: Spectral weight analysis. (a1)} Spectra of FTS along $\Gamma$-$M$ direction at $h\nu=26$ eV. \textbf{(a2)} Energy distribution curves (EDCs) along cut C1 and C2 in (a1). \textbf{(a3)} EDC along cut C3 in (a1). \textbf{(b1)-(b3)} similar to (a1)-(a3), with FTS along $\Gamma$-$M$ direction at $h\nu=44$ eV. \textbf{(c1)-(c3)} similar to (a1)-(a3), collected on FeSe along $\Gamma$-$M$ direction at $h\nu=26$ eV. The EDCs in (c3) is averaged in a finite momentum window marked by two orange lines in (c1). \textbf{(d1)-(d3)} similar to (c1)-(c3), collected on FeSe along $\Gamma$-$M$ direction at $h\nu=44$ eV. All spectra are taken under $p$ polarization. Energy averaging windows along energy axis are marked by grey area.}
\end{figure*}

\begin{figure*}[h]%
\centering
\includegraphics[scale=0.8]{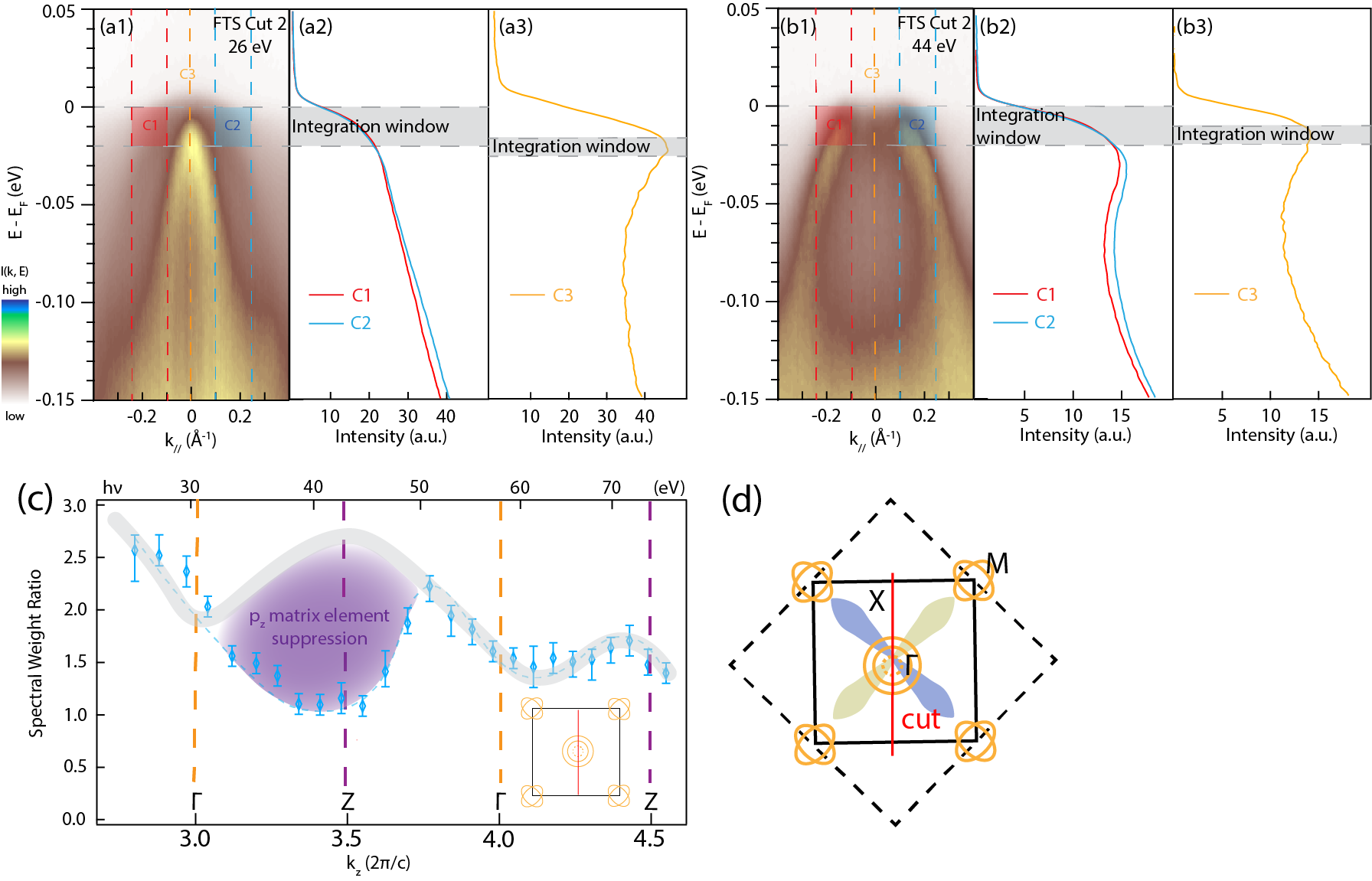}
\caption{\textbf{Fig. S5: Spectral weight analysis for FTS along $\Gamma$-$X$ direction. (a1)} Spectrum of FTS along $\Gamma$-$X$ direction at $h\nu=26$ eV. \textbf{(a2)} EDCs along C1 and C2 in (a1). \textbf{(a3)} EDC along C3 in (a1). \textbf{(b1)-(b3)} similar to (a1)-(a3), with FTS along $\Gamma$-$X$ direction at $h\nu=44$ eV. \textbf{(c)} Spectral weight ratio as a function of $k_z$ for FTS along $\Gamma$-$X$ direction. Purple shaded area marks the anomalous matrix element depression. Inset shows cut orientation. \textbf{(d)} Schematic of the cut geometry along $\Gamma$-$X$ direction similar to Fig. 1(b). The $d_{xz}$ and $d_{yz}$ orbitals are both 45$^\circ$ offset from the cut plane, and therefore have similar matrix elements by symmetry.}
\end{figure*}

\begin{figure*}[h]%
\centering
\includegraphics[scale=0.8]{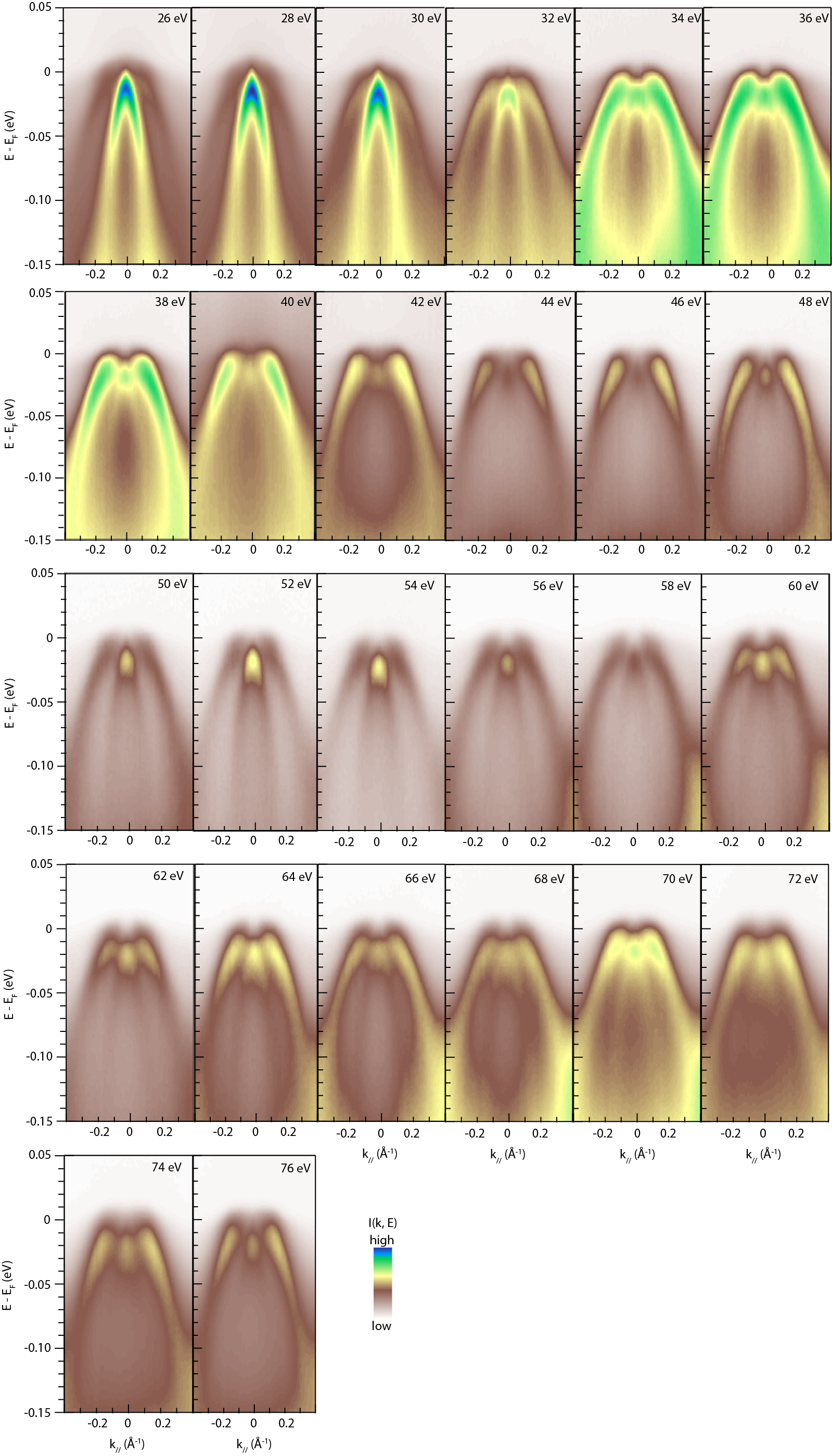}
\caption{\textbf{Fig. S6: FTS spectra taken along $\Gamma$-$M$ direction at 26-76 eV under $p$ polarization.}}
\end{figure*}

\begin{figure*}[h]%
\centering
\includegraphics[scale=0.8]{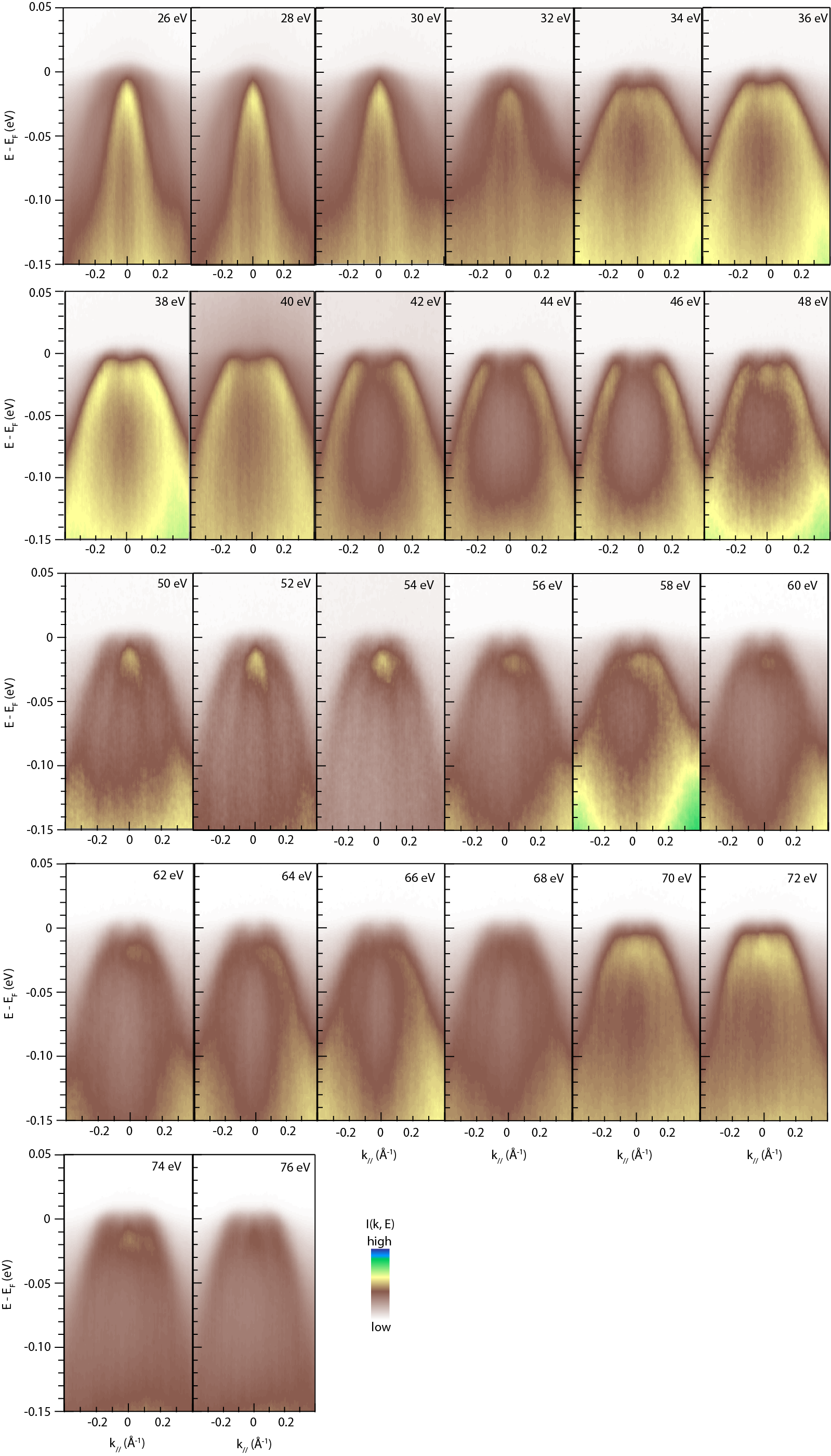}
\caption{\textbf{Fig. S7: FTS spectra taken along along $\Gamma$-$X$ direction at 26-76 eV under $p$ polarization.}}
\end{figure*}

\begin{figure*}[h]%
\centering
\includegraphics[scale=0.8]{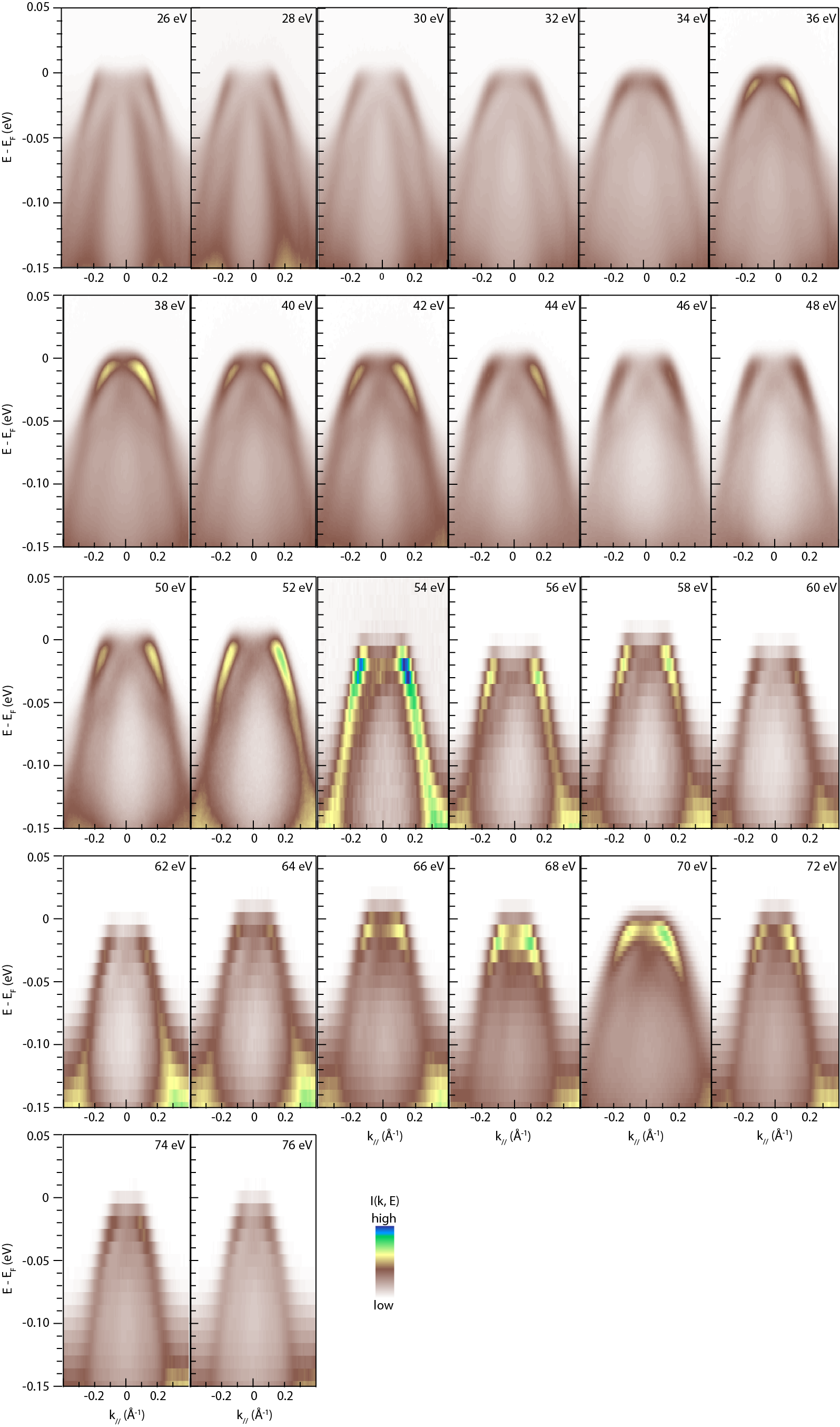}
\caption{\textbf{Fig. S8: FeSe spectra taken along $\Gamma$-$M$ direction at 26-76 eV under $p$ polarization.}}
\end{figure*}

\section{Matrix element calculation}
\subsection{Formulation}
In the following, we show details on calculating the matrix elements\cite{moser2017experimentalist, wang2012orbital, day2019computational, goldberg1981photoionization, Hufner}. Under the dipole approximations, the photoemission matrix element is \cite{moser2017experimentalist,day2019computational, Hufner}
\begin{equation*}
M \propto  \bra{f}\hat{\boldsymbol\epsilon}\cdot \boldsymbol r\ket{i}
\end{equation*}
, where $\ket{f}$ is the final state, $\ket{i}$ the initial state, $\hat{\boldsymbol \epsilon}$ unit vector of electric field, and $\boldsymbol r$ spatial vector in solids.

We use a free-electron plane-wave state to approximate the final state and expand it as well as $\hat{\boldsymbol\epsilon}\cdot \boldsymbol r$ in the basis of spherical harmonics.
\begin{equation*}
    \braket{\boldsymbol r}{f} = e^{i\boldsymbol k \boldsymbol r} = 4\pi\sum_{l_f=0}^{\infty}i^{l_f}j_{l_f}(\boldsymbol k_f \boldsymbol r)\sum_{m_f = -l_f}^{l_f}\bar{Y}_{l_f}^{m_f}(\theta_k,\phi_k)Y_{l_f}^{m_f}(\theta,\phi)
\end{equation*}
\begin{equation*}
  \hat{\boldsymbol \epsilon}\cdot \boldsymbol r = \frac{\boldsymbol r}{3}\sum_{m_\epsilon = -1}^{+1}\bar{Y}^{m_\epsilon}_1(\theta_\epsilon,\phi_\epsilon)Y_1^{m_\epsilon}(\theta,\phi)
\end{equation*}
where $j_{l_f}(\boldsymbol k_f \boldsymbol r)$ is the Bessel function, and $Y(\bar{Y})$ is the spherical harmonic function (complex conjugate of spherical harmonic function).

We also approximate the initial state using hydrogen-like orbitals:
\begin{equation*}
\braket{\boldsymbol r}{i} = R_{nl}(r)Y_l^m(\theta, \phi)
\end{equation*}
\begin{equation*}
    R_{nl} = \sqrt{(\frac{2}{nr_B})^3\frac{(n-l-1)!}{2n(n+1)!}}(\frac{2Z r}{nr_B})^le^{-\frac{Z r}{nr_B}}L_{n-l-1}^{2l+1}(\frac{2Zr}{nr_B})
\end{equation*}
, where $R_{nl}(r)$ the radial component, $Y_l^m(\theta, \phi)$ the spherical harmonic, $r_B$ the Bohr radius, $Z$ the effective nuclear charge (determined by Slater's rule, $Z = 6.25$ for Fe $3d$ and $Z=6.95$ for Se $4p$ and Te $5p$), $L_{n-l+1}^{2l+1}$ the generalized Laguerre polynomial.

Putting together,
\begin{widetext}
\begin{equation}
\begin{split}
M_{nlm} \propto \frac{4\pi}{3}\sum_{l_f=0}^{\infty}i^{l_f}\int_{0}^{\infty}dr r^3R_{nl}(r)j_{l_f}(k_fr)\sum_{m_\epsilon=-1}^{1}\sum_{m_f=-l_f}^{l_f}\bar{Y}_1^{m_\epsilon}(\theta_\epsilon, \phi_\epsilon)\bar{Y}_{l_f}^{m_f}(\theta_k,\phi_k)\\\int_{0}^{\pi}d\theta\sin\theta\int_{0}^{2\pi}d\phi Y_l^m(\theta, \phi)Y_1^{m_\epsilon}(\theta, \phi)Y_{l_f}^{m_f}(\theta,\phi)
\end{split}
\end{equation}
\end{widetext}
where $n,l,m$ index the atomic wave function as the initial state. The integral will only be non-zero if
\begin{equation*}
\begin{cases}
    l_f - l = \pm 1\\
    m_f = -(m+m_\epsilon)    
\end{cases}
\end{equation*}
We can therefore further simplify eq. (5) to 
\begin{widetext}
\begin{equation}
\begin{split}
    M_{nlm}\propto \frac{4\pi}{3}\sum_{l_f=l\pm 1}\sum_{m_\epsilon=-1}^{+1}i^{l_f}I_R(n,l,l_f, k_f){(-1)}^{m_\epsilon}{(-1)}^{-(m+m_\epsilon)}\\Y_1^{-m_\epsilon}(\theta_\epsilon,\ \varphi_\epsilon)Y_{l_f}^{m+m_\epsilon}(\theta_k,\ \varphi_k)I_\Omega(l,m,l_f,-(m+m_\epsilon),m_\epsilon)
\end{split}
\end{equation}    
\end{widetext}
, where 
\begin{equation}\label{eq: radial integral approximation 1}
I_R\left(n,l,l_f,k_f\right)=\int_{0}^{\infty}{dr\ r^3R_{nl}(r)j_{l_f}(k_fr)}
\end{equation}
We can further represent eq. (6) in complex spherical harmonics 
\begin{widetext}
\begin{equation}
    M_{nlm}(\theta_\epsilon,\varphi_\epsilon,\theta_k,\varphi_k,k_f) =  
    \begin{cases}
    \frac{i}{\sqrt{2}}(M_{nlm} - (-1)^mM_{nl(-m)}) & m>0\\
    M_{nl0}, & m=0\\
    \frac{i}{\sqrt{2}}(M_{nl(-m)} + (-1)^mM_{nlm}) & m<0\\
    \end{cases}
\end{equation}    
\end{widetext}
by transformation from the complex basis of spherical harmonics $Y_l^m: S^2 \rightarrow \mathbb{C}$ to the real basis of spherical harmonics $Y_{lm}:S^2\rightarrow \mathbb{R}$:
\begin{equation*}
    Y_{lm} =  
    \begin{cases}
    \frac{i}{\sqrt{2}}(Y_l^m - (-1)^mY_l^{-m}), & m>0\\
    Y_l^0, & m=0\\
    \frac{i}{\sqrt{2}}(Y_l^{-m} + (-1)^mY_l^m), & m>0
    \end{cases}       
\end{equation*}

\subsection{Parameters from experimental setting}
At the Stanford synchrotron radiation lightsource (SSRL) beamline station 5-2, for $p$ polarization, $\theta_\epsilon={40}^\circ$ and $\phi_\epsilon={90}^\circ$; for $s$ polarization, $\theta_\epsilon={90}^\circ$ and $\phi_\epsilon=0^\circ$, with the definition of the coordinate system shown in Fig. S9. To calculate the matrix element at different momentum points, $(\theta_k,\phi_k)$ are transformed from $(k_x,k_y)$ by mapping between photoemission angle and the in-plane momentum\cite{ishida2018}.
\begin{figure*}[h]%
\centering
\includegraphics[scale=0.8]{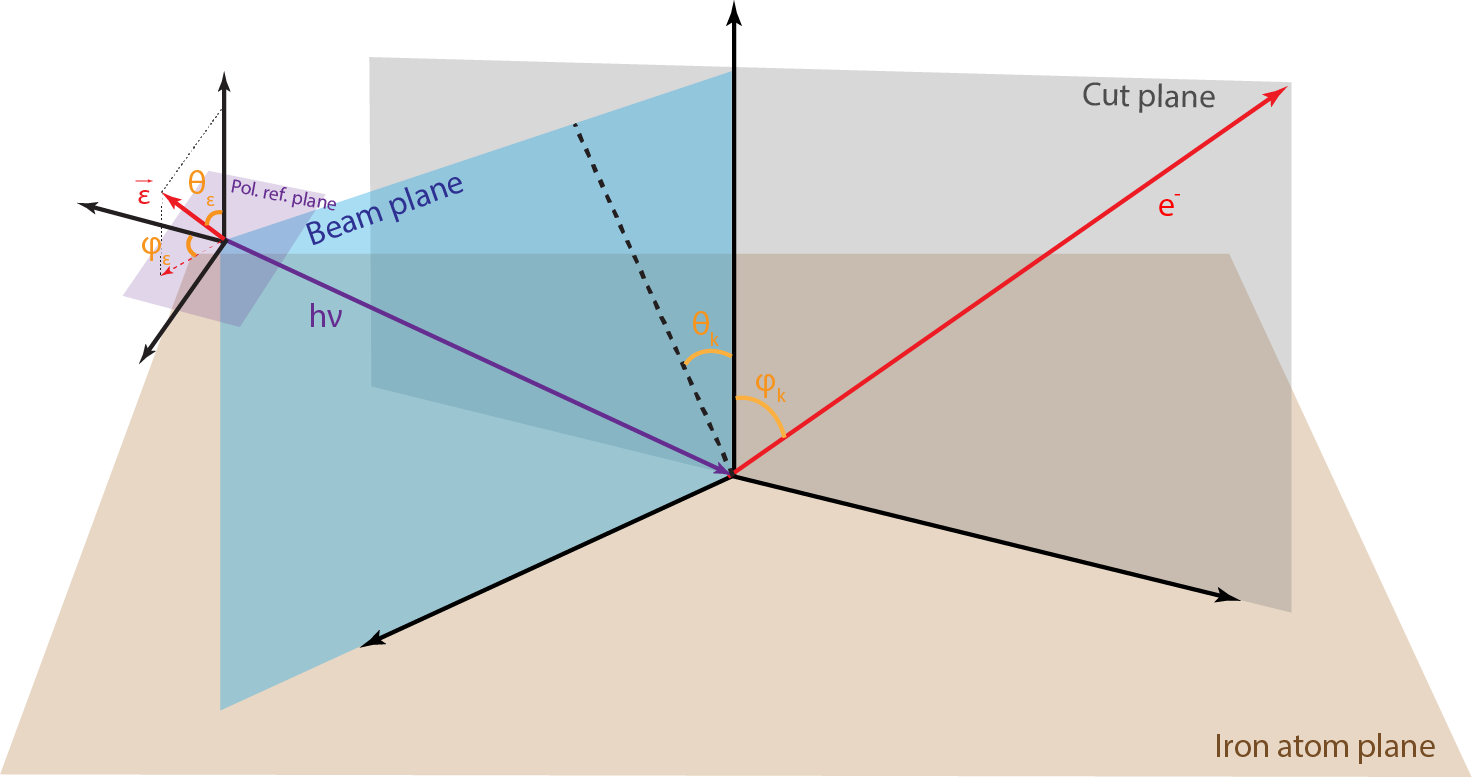}
\caption{\textbf{Fig. S9: Schematic of the definition of the coordinate system in the matrix element calculation.} Brown plane: iron-atom plane; grey plane: photoemission (cut) plane; blue plane: beam plane; purple plane: polarization reference plane.}
\end{figure*}
\subsection{Photon energy dependence and in-plane momentum dependence of matrix elements }

In Fig. S11(a-c) we show the matrix element for $4p_z$, $5p_z$, $d_{yz}$ and $d_{xz}$ orbitals up to a normalization constant as a function of photon energy at different in-plane momentum. In Fig. S10(d-f) we show the relative matrix element of $d_{xz}$ and $p_z$ normalized by that of $d_{yz}$. Fig. S11 shows the normalized matrix element ratio after equal-weight mixing wave functions for $4p_z$ and $5p_z$ atomic orbitals and averaging over a momentum window which persistently covers the spectral weight analysis area (red and orange bars in Fig. 3(a)) at all photon energies. Fig. S12 show calculated matrix elements under $s$ polarization.

\begin{figure*}[h]%
\centering
\includegraphics[scale=0.8]{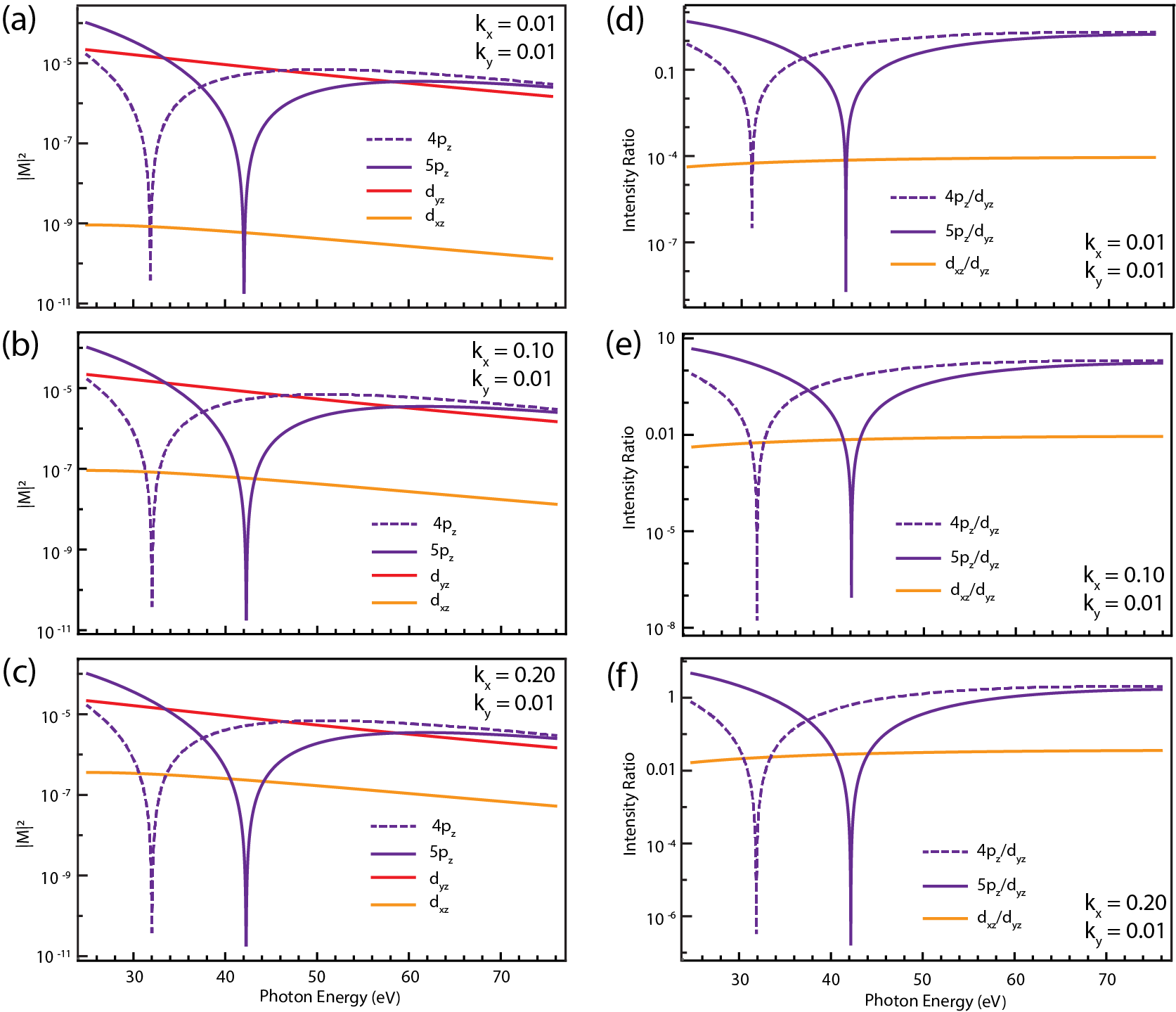}
\caption{\textbf{Fig. S10: Matrix element calculation under $p$ polarization. (a-c)} Matrix element for $4p_z$, $5p_z$, $d_{yz}$ and $d_{xz}$ orbitals at (a) $k_x=0.01$, (b) $k_x=0.1$, (c) $k_x=0.2$. \textbf{(b-d)} Relative matrix element of $d_{xz}$ and $p_z$ normalized by that of $d_{yz}$ from (a-c).}
\end{figure*}

\begin{figure*}[h]%
\centering
\includegraphics[scale=0.8]{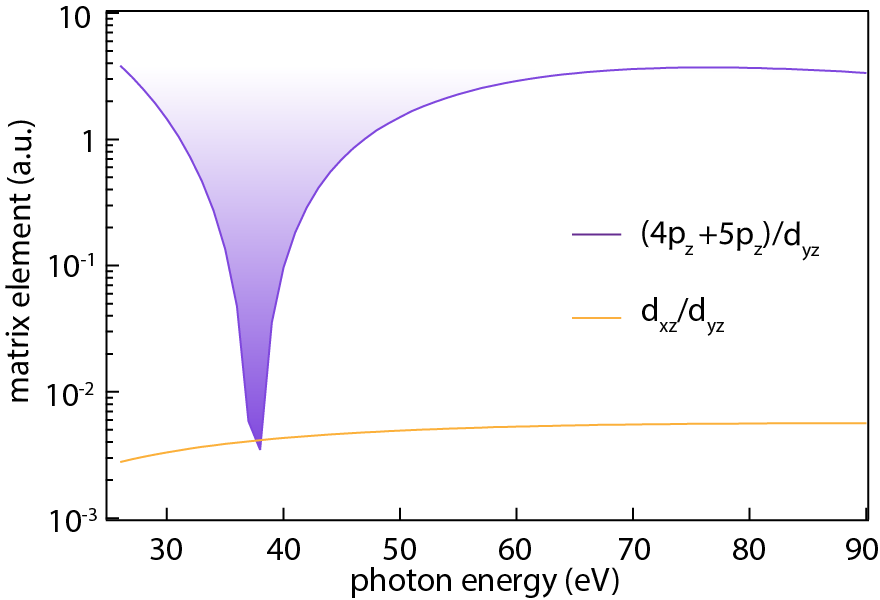}
\caption{\textbf{Fig. S11: matrix element of Se/Te $p_z$ and Fe $d_{xz}$ orbitals normalized by Fe $d_{yz}$ orbitals.}}
\end{figure*}

\begin{figure*}[h]%
\centering
\includegraphics[scale=0.8]{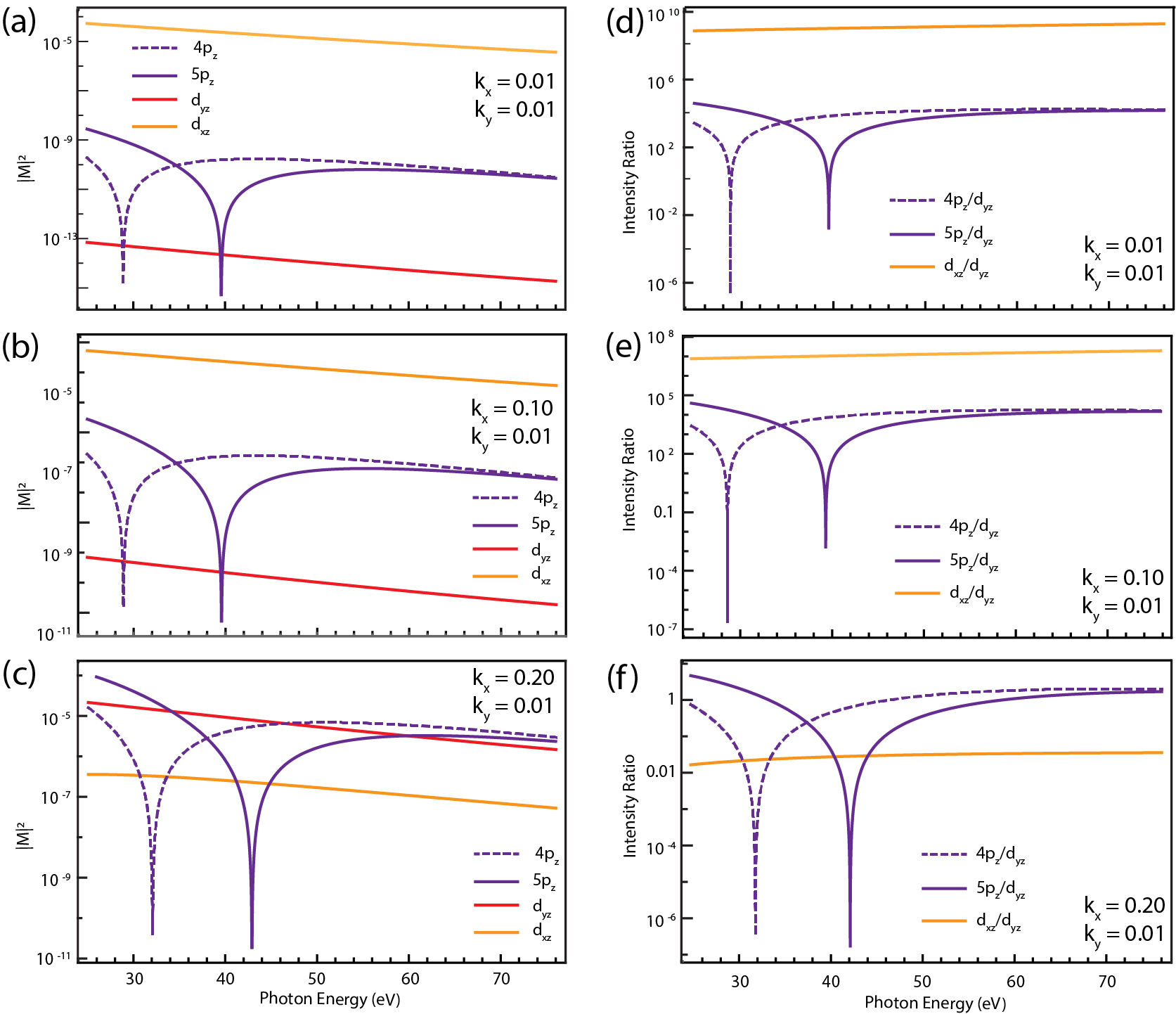}
\caption{\textbf{Fig. S12: Photon energy dependence of the matrix element calculation under $s$ polarization. (a-c)} Matrix element for $4p_z$, $5p_z$, $d_{yz}$ and $d_{xz}$ orbitals at (a) $k_x=0.01$, (b) $k_x=0.1$, (c) $k_x=0.2$. \textbf{(b-d)} Relative matrix element of $d_{xz}$ and $p_z$ normalized by that of $d_{yz}$ from (a-c).}
\end{figure*}

In Fig. S13, we show the matrix element for $4p_z$, $5p_z$, $d_{yz}$, and $d_{xz}$ orbitals up to a normalization constant as a function of in-plane momentum along $\Gamma$-$M$ at $h\nu=26$ eV. The matrix element at $k_\parallel=0$ is zero for $d_{xz}$ orbitals and finite for the other orbitals.
\begin{figure*}[h]%
\centering
\includegraphics[scale=0.8]{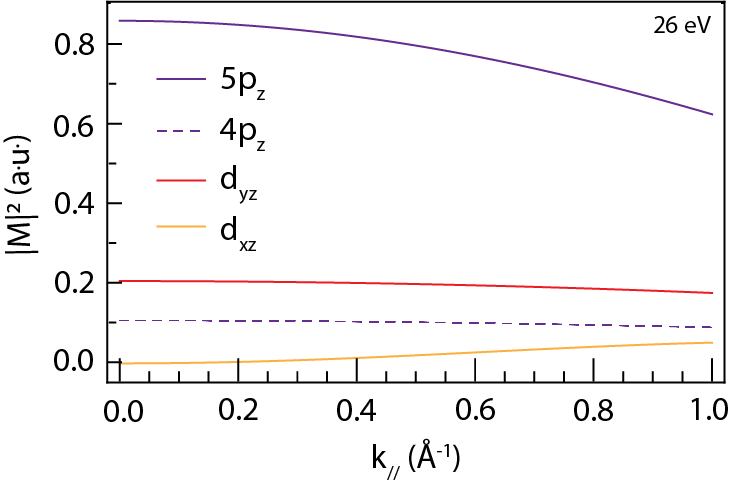}
\caption{\textbf{Fig. S13 In-plane momentum dependence of matrix element calculation with $p$ polarization at $h\nu=26$ eV.}}
\end{figure*}

\subsection{Formulation under alternative approximations}
The wave function of initial and final states can be alternatively approximated by the bound states and scattering states under Hartree-Fock central potential, respectively\cite{yeh1985atomic, goldberg1981photoionization}. The wave function of initial state can generally be decomposed as a product of radial part and angular part,
\begin{widetext}
\begin{equation*}
    \braket{\boldsymbol{r}|\Phi_{nlx}} = R_{n,l}(r) \sum_{m=-l}^l n(m) Y_{l,m}(\theta,\phi) := \frac{P_{n,l}(r)}{r} \sum_{m=-l}^l n(m) Y_{l,m}(\theta,\phi) 
\end{equation*}
\end{widetext}
, where $R_{n,l}(r)$ is the radial part of the wave function with normalization condition:
\begin{equation*}
    \int_0^\infty P_{n,l}^2(r) dr = 1
\end{equation*}, $P_{n,l}(r):= rR_{n,l}(r)$ , $Y_{l,m}(\theta,\phi)$ is the complex spherical harmonics. By convention, the initial atomic orbital states are indexed by $s$, $p_x$, $p_y$, $p_z$, $d_{xy}$, $d_{xz}$, $d_{yz}$, $d_{x^2-y^2}$, $d_{z^2}$, etc., which directly corresponds to real spherical harmonics. The $n(m)$ is the coefficient which transform real spherical harmonics $Y_{l,m}^{(R)}(\theta, \phi)$ into the complex spherical harmonics $Y_{l,m}(\theta, \phi)$ basis by the relation:
\begin{equation*}
    Y_{l,m}^{(R)} = \begin{cases}
        \frac{i}{\sqrt{2}}(Y_{l,m} - (-1)^mY_{l, -m}), \quad \text{if } m<0\\
        Y_{l,0}, \quad \text{if } m =0\\
        \frac{1}{\sqrt{2}}(Y_{l,-m} + (-1)^mY_{l,m}), \quad \text{if } m>0
    \end{cases}
\end{equation*}
The radial part of wave function $P_{n,l}(r):= \frac{R_{n,l}(r)}{r}$ are obtained by solving the one-electron Schrodinger equation
\begin{equation*}
    [\frac{d^2}{dr^2} + V(r) + \epsilon_{n,l} - \frac{l(l+1)}{r^2}] P_{n,l}(r) = 0
\end{equation*}
, where $V(r)$ is the effective central-field potential, and $\epsilon_{n,l}$ is the binding energy. Here, we take the form
\begin{equation*}
    V(r) = V^H(r) + V^{\text{ex.}}(r)
\end{equation*}
, where $V^H(r)$ is the standard Hartree potential and $V^{\text{ex.}}(r)$ is the free-electron exchange potential in the Slater form:
\begin{equation*}
    V^{\text{ex.}}(r) = -6[(\frac{3}{8\pi})\cdot \rho(r)]^{1/3}
\end{equation*}
, where $\rho(r)$ is the charge density. The asymptotic behavior of $V(r)$ satisfies:
\begin{equation*}
    \lim_{r\rightarrow 0} V(r) = \frac{2Z}{r}; \quad \lim_{r\rightarrow \infty} V(r) = \frac{2}{r}
\end{equation*}
, where $Z$ is the nuclear charge.

The wave function of the final state is taken as the continuum (scattering) state of one-electron wave function under the same potential and can be written as a partial-wave expansion
\begin{widetext}
\begin{equation*}
    \braket{\boldsymbol{r}|\Phi_{E_{\text{kin}, \boldsymbol{k}}}} = 4\pi \sum_{l',m'} (i)^{l'} \exp(-i\delta_{l'}) Y_{l',m'}^*(\theta_k,\phi_k) Y_{l',m'}(\theta,\phi) R_{E_\text{kin},l'} (r)
\end{equation*}    
\end{widetext}
, where $\theta_k, \phi_k$ is the direction of the photoemited electron in spherical coordinate system  and $Y_{l,m}^*$ is the complex conjugate of $Y_{lm}$ with relation $ Y_{l,m}^*= (-1)^mY_{l,-m}$. $R_{E_{\text{kin}}, l'}(r)$ (equivalently, $P_{E_{\text{kin}}, l'}/r$) is the radial part of one-electron wave function under the same effective central potential obtained by solving one-electron Schrodinger equation:
\begin{widetext}
\begin{equation*}
    [\frac{d^2}{dr^2} + V(r) + E_{\text{kin}} - \frac{l(l+1)}{r^2}] P_{E_{\text{kin}},l}(r) = 0
\end{equation*}
, with $E_{\text{kin}} = h\nu - \epsilon_{nl}$. $P_{E_{\text{kin}},l}(r)$ can be normalized by the asymptotic behavior at infinity:
\begin{equation*}
    \lim_{r\rightarrow \infty} P_{\epsilon, l}(r) = \pi^{-1/2} \epsilon^{-1/4} \sin [\epsilon^{1/2}-\frac{1}{2}l\pi - \epsilon^{-1/2}\log (2\epsilon^{1/2}r) + \delta_l(\epsilon)]
\end{equation*}    
\end{widetext}
, where $\delta_l(\epsilon)$ is the phase shift.

The dipole operator the $\hat{\boldsymbol{\varepsilon}}\cdot \boldsymbol{r}$ are expressed in spherical coordinate system by
\begin{equation*}
    \begin{cases}
    \varepsilon_x = \sin\theta_\varepsilon \cos\phi_\varepsilon\\
    \varepsilon_y = \sin\theta_\varepsilon \sin\phi_\varepsilon\\
    \varepsilon_z = \cos\theta_\varepsilon
    \end{cases}
\end{equation*}
and 
\begin{equation*}
    \begin{cases}
    x/r = \sin\theta \cos\phi = (2\pi/3)^{1/2} (-Y_{1,1} + Y_{1,-1})\\
    y/r = \sin\theta \sin\phi = i(2\pi/3)^{1/2} (Y_{1,1} + Y_{1,-1})\\
    z/r = \cos\theta = (4\pi/3)^{1/2}Y_{1,0}
    \end{cases}
\end{equation*}
Putting altogether, we have 
\begin{widetext}
\begin{equation*}
\begin{split}
    \bra{\Phi_{E_{\text{kin}, \boldsymbol{k}}}}\hat{\boldsymbol{\varepsilon}}\cdot \boldsymbol{r}\ket{\Phi_{nlx}} = 4(2/3)^{1/2}\pi^{3/2} \sum_{l',m',m} n(m) (-i)^{l'} \exp(i\delta_{l'}) Y_{l',m'}(\theta_k, \phi_k)R_{n,l'}(E_\text{kin})\\
    \times [\varepsilon_x (-\bra{l',m'}1,1\ket{l,m} + \bra{l'm'}1,-1\ket{l,m}) + i\varepsilon_y (\bra{l',m'}1,1\ket{l,m} + \bra{l',m'}1,-1\ket{l,m})\\ + \sqrt{2}\varepsilon_z \bra{l',m'}1,0\ket{l,m}]
\end{split}
\end{equation*}
\end{widetext}
, where 
\begin{equation}\label{eq: radial integral approximation 2}
    R_{n, l\pm1}(\epsilon) = \int_0^\infty P_{n,l}(r)rP_{\epsilon,l\pm1}(r)dr
\end{equation}
\begin{widetext}
\begin{equation}
\begin{split}
    \bra{l,m}\hat{l}, \hat{m}\ket{l',m'} := \int Y^*_{l,m}(\theta, \phi)Y_{\hat{l},\hat{m}}(\theta, \phi)Y_{l',m'}(\theta, \phi)d\Omega\\
    = \int (-1)^m Y_{l, -m}(\theta, \phi) Y_{\hat{l}, \hat{m}}(\theta, \phi) Y_{l', m'}(\theta, \phi) d\Omega \\
    =: (-1)^m \sqrt{\frac{3(2l+1)(2l_f+1)}{4\pi}}
     \begin{pmatrix}l & \hat{l} & l'\\0 & 0 & 0\end{pmatrix}
     \begin{pmatrix}l & \hat{l} & l'\\-m & \hat{m} & m'\end{pmatrix}
\end{split}
\end{equation}.    
\end{widetext}
, where in the last line we use the definition of Wigner 3-jm symbols. Specifically, for $\hat{l}=1$, $\bra{l,m}1, \hat{m}\ket{l',m'}$ is only non-zero when
\begin{equation*}
    \begin{cases}
    l \pm 1 = l' \\
    m + \hat{m} + m'= 0
    \end{cases}
\end{equation*}
, which recovers the dipole selection rule again.

\subsection{Comparison under different approximations}
The photon energy dependence of matrix elements under the alternative approximations are shown in Fig. S14, where no matrix element suppression for $p_z$ is observed throughout the phonon energy range considered. In table S2, we summarize the existence of matrix element suppression under different approximations. A closer look of the calculation suggests that the matrix element suppression arises from the diminishing radial integral of the wave function in eq. \ref{eq: radial integral approximation 1} and eq. \ref{eq: radial integral approximation 2}, which highly depends on the details of wave functions. Therefore, we treat with care the speculation of the existence of matrix element suppression in the main text discussion.

\begin{figure*}[h]%
\centering
\includegraphics[scale=0.8]{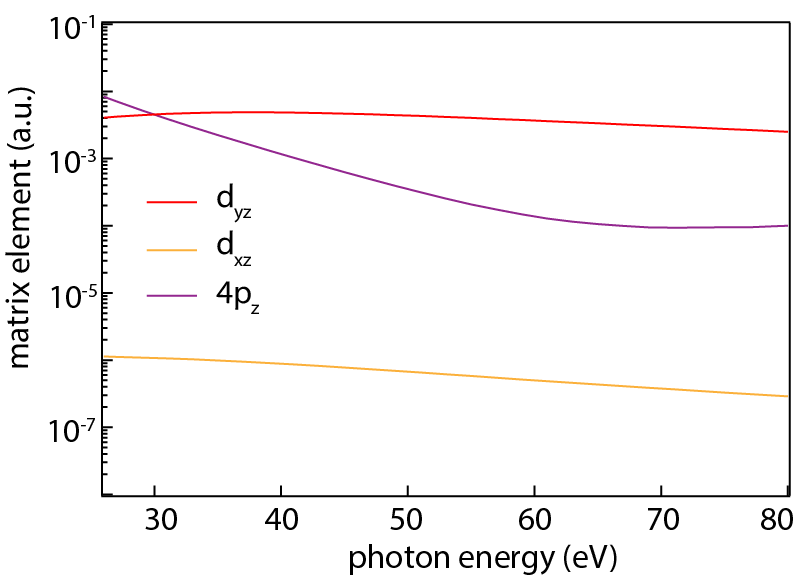}
\caption{\textbf{Fig. S14: matrix element of Se/Te $p_z$ and Fe $d_{xz}$ orbitals under alternative approximations.}}
\end{figure*}

\begin{table*}[h]
\begin{center}
\caption{TABLE S2. Matrix element suppression under different approximations}
\begin{tabular}{p{6cm}||c}
\toprule
approximation configurations & existence of matrix element suppression\\
\hline
hydrogen-like atomic orbital + free-electron & Y \\ 
bound states under HF potential + scatter state under HF potential & Y\\
hydrogen-like atomic orbital + scatter state under HF potential & N\\
bound states under HF potential + scatter state under HF potential & N\\
\botrule
\end{tabular}
\end{center}
\end{table*}

\color{black}
\section{DFT calculations and tight-binding model}
\subsection{DFT calculations and symmetry analysis}

The \textit{density functional theory} (DFT) calculations were performed using the open-source software ABINIT\cite{abinit}, with the experimental FeSe structure\cite{structure} as starting point (see table S4). \textit{Spin-orbit coupling} (SOC) was included in all DFT calculations. 
The doping was modelled within the \textit{virtual crystal approximation}\cite{vca} (VCA) available in ABINIT, which requires the use of norm-conserving pseudopotentials\cite{oncvpsp} accessible in the \textit{pseudodojo} repository\cite{pseudodojo}.

\begin{table*}[h]
\begin{center}
\caption{TABLE S3. Matrix element suppression under different approximations}
\begin{tabular}{p{6cm}||c}
\toprule
approximation configurations & existence of matrix element suppression\\
\hline
hydrogen-like atomic orbital + free-electron & Y \\ 
bound states under HF potential + scatter state under HF potential & Y\\
hydrogen-like atomic orbital + scatter state under HF potential & N\\
bound states under HF potential + scatter state under HF potential & N\\
\botrule
\end{tabular}
\end{center}
\end{table*}

\begin{table*}[h!]
\begin{center}
\caption{TABLE S4. Dimensions of the tetragonal unit cell and positions of the atoms.}
	\begin{tabular}{cccc}
		\hline \hline
		\multicolumn{4}{c}{Lattice vectors}                   \\ \hline
		Vector    & $a_1$        & $a_2$        & $a_3$       \\
		Length (Ang.)    & $3.50445975$ & $3.50445975$ & $5.5614843$ \\ \hline
		\multicolumn{4}{c}{Atomic positions (fractional)}     \\ \hline
		Fe (1)    & $0.0$        & $0.0$        & $0.0$       \\
		Fe (2)    & $0.5$        & $0.5$        & $0.0$       \\
		Te/Se (1) & $0.0$        & $0.5$        & $0.764815$  \\
		Te/Se (2) & $0.5$        & $0.0$        & $0.235185$ \\
        \hline \hline
	\end{tabular}
\end{center}
\label{tab:crystal_str}
\end{table*}

The self-consistent cycle was performed on a $17 \times 17 \times 9$ grid with a plane-wave cut-off energy of $80$ Hartree. Afterwards, the band structure [see Fig. S16] was calculated non-self-consistently with at least $200$ points in the smallest segment of the path.

\begin{figure*}[h]%
\centering
\includegraphics[scale=0.75]{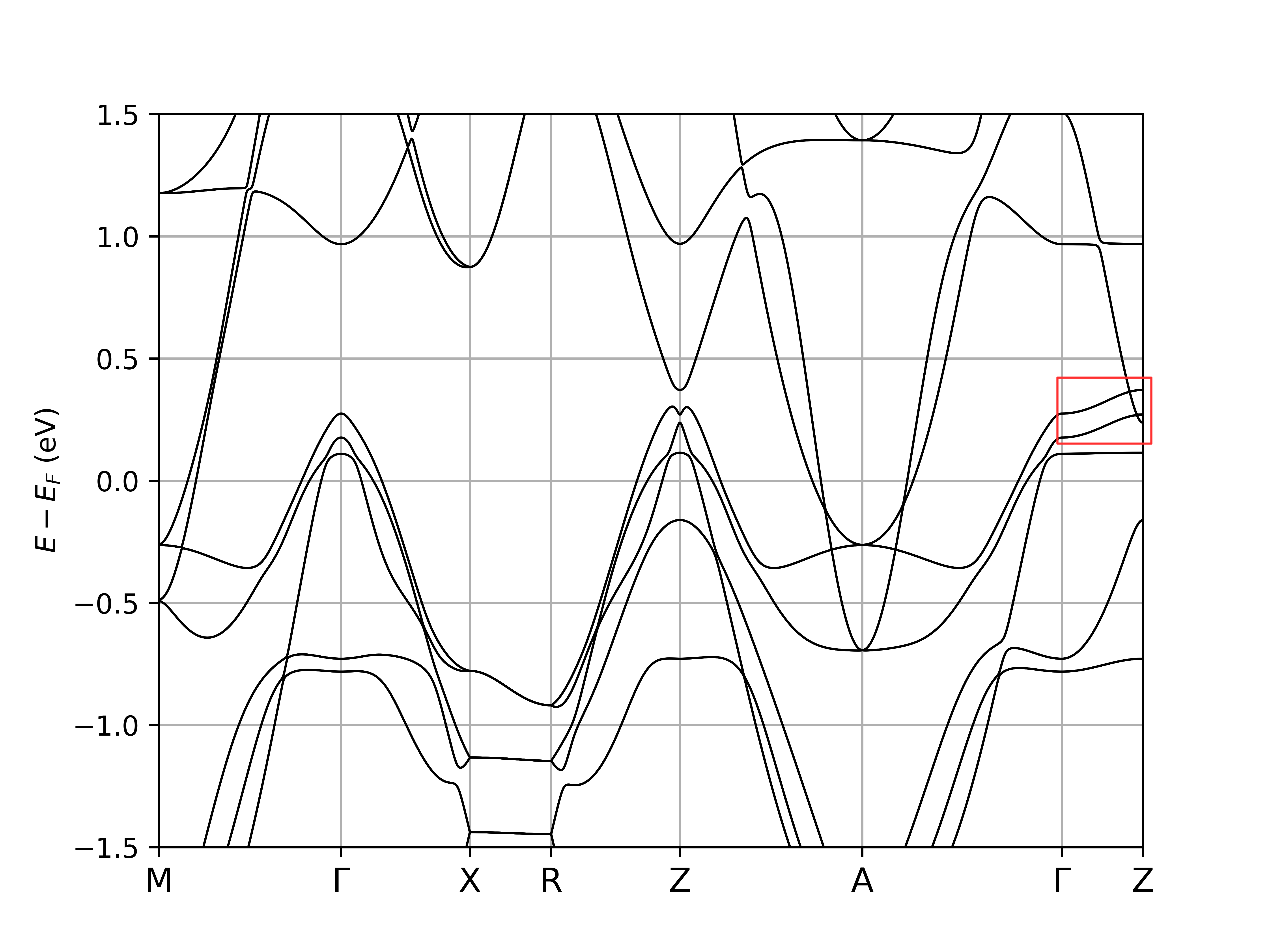}
\caption{\textbf{Fig. S15: Band structure of the FeTe\textsubscript{0.55}Se\textsubscript{0.45}. The red square marks the area of interest in this work.}}
\label{fig: bs}
\end{figure*}

Fe atoms sit at the Wyckoff position (WP) $2a$ and contribute to the states around the Fermi level mainly with $d$-type orbitals.
The site-symmetry group (SSG) of sites in WP $2a$ is isomorphic to $\bar4m2$.
It can be shown that the pair $(d_{xz},d_{yz})$ cannot be separated into different subspaces due to this symmetry -- for instance, the 4-fold rotation turns $d_{xz}$ into $d_{yz}$, and vice-versa.
Therefore, this pair of orbitals transforms as a sum of irreducible representations (irreps) of the SSG which we write symbolically as

\begin{equation}
        \left[(d_{xz},d_{yz})\times (\Uparrow,\Downarrow)\right] \downarrow G_{2a} = \bar{E}_1 \oplus \bar{E}_2.
\end{equation}
Here, the pair $(\Uparrow,\Downarrow)$ indicates the two spin orientations, while $\downarrow$ alone is the subduction operation defined within the context of representations of symmetry groups.
According to the theory of topological quantum chemistry\cite{tqc} (TQC), these orbitals induce a band representation with the following irreps on the line $\Delta$ connecting $\Gamma$ to $Z$:

\begin{equation} \label{eq:d_irreps}
\begin{split}
    & (\bar{E}_1 \uparrow G)_{2a}: \{ \bar\Gamma_7 \oplus \bar\Gamma_8, \bar Z_7 \oplus \bar Z_8, \bar\Lambda_6 \oplus \bar\Lambda_7 \}, \\
    & (\bar{E}_2 \uparrow G)_{2a}: \{ \bar\Gamma_6 \oplus \bar\Gamma_9, \bar Z_6 \oplus \bar Z_9, \bar\Lambda_6 \oplus \bar\Lambda_7 \}.
\end{split}
\end{equation} 

Te/Se atoms sit at WP $2c$, and contribute to the bands around the Fermi level with $p_z$ orbitals.
These spinful orbitals belong to an irrep of the SSG at WP $2c$ on their own:

\begin{equation}
       \left[p_z \times (\Uparrow,\Downarrow)\right] \downarrow G_{2c} = \bar{E}_1.
\end{equation}

According to the TQC formalism, the little-group irreps of this band representation at $\Gamma$, $Z$, and $\Lambda$ are:

\begin{equation}\label{eq:p_irreps}
    ( \bar{E}_1 \uparrow G )_{2c}: \{ \bar\Gamma_7 \oplus \bar\Gamma_9, \bar Z_7 \oplus \bar Z_9, 2 \bar\Lambda_7 \}.
\end{equation}

\begin{figure*}[h]%
\centering
\includegraphics[scale=0.6]{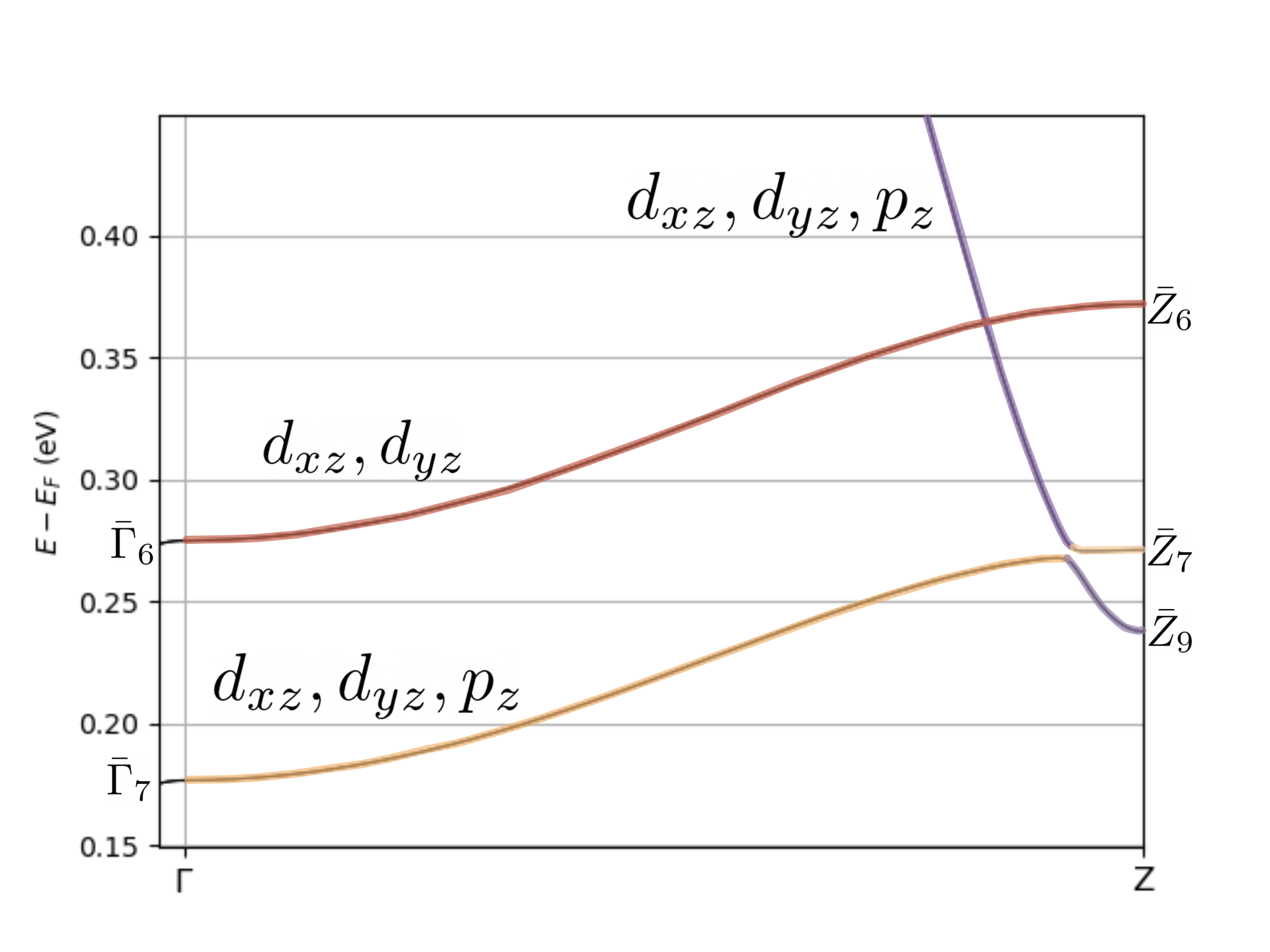}
\caption{\textbf{Fig. S16 Zoom of the and structure of the FeTe\textsubscript{0.55}Se\textsubscript{0.45} around the Fermi level along the line $\Lambda$ joining $\Gamma$ and $Z$. The color codes indicate the dominant orbital character of irreps under the Wannier representation, consistent with Fig. 1(e) and Fig. 4.}}
\label{fig:gm_z_irreps}
\end{figure*}

Fig.S16 shows the orbital content of the bands, which is compatible with the symmetry properties at these two points. In particular, cyan bands carry the irrep $\bar \Lambda_7$, and can be either $d$ or $p$ orbitals, or a mix of the two, since this irrep is present in the band representations of both kinds of orbitals [see Eqs.~\eqref{eq:d_irreps} and \eqref{eq:p_irreps}].
In contrast, the orange band has $d$ character only, as it carries the irrep $\bar \Lambda_6$ stemming exclusively from the band representation of $d$-orbitals in Eq.~\eqref{eq:d_irreps}. Furthermore, the analysis shows that the gap can be due to a hybridisation between the two kinds.

\subsection{Tight-binding model}

Based on the irreps in Eqs.~\eqref{eq:d_irreps} and \eqref{eq:p_irreps}, we build a general model that respects the full symmetry of the crystal. The coupling terms between the basis elements are added in order of ascending distance to show the full set of features observed experimentally. The irrep basis can be related to the conventional $d$ and $p$ orbital basis by the transformations

\begin{equation} \label{eq:bases}
\begin{split}
    a_{i1} = \frac{1}{\sqrt{2}}[d_{xz}^\downarrow - i d_{yz}^\downarrow] &, \quad a_{i2} = \frac{e^{-i\pi/4}}{\sqrt{2}}[d_{xz}^\uparrow + i d_{yz}^\uparrow] \\
    b_{i1} = \frac{1}{\sqrt{2}}[d_{xz}^\downarrow + i d_{yz}^\downarrow] &, \quad b_{i2} = \frac{e^{3i\pi/4}}{\sqrt{2}}[d_{xz}^\uparrow - i d_{yz}^\uparrow] \\
    c_{i1} = p_z^\uparrow &,\quad c_{i2} = e^{-i\pi/4}p_z^\downarrow.
\end{split}
\end{equation}

where the states have been labelled in the following way:

\begin{itemize}
    \item $a_{ij}$ is the $j$-th orbitals of $\bar{E}_2$ at position $i$ of the WP $2a$. Here $i=1,2$ and $j=1,2$ because the WP $2a$ contains two sites -- \textit{i.e.} has multiplicity two -- and $\bar{E}_2$ is two-dimensional.
    \item $b_{ij}$ follows the same convention as $a_{ij}$, but for the irrep $\bar{E}_1$.
    \item $c_{ij}$ follows the same convention as $a_{ij}$, but for the irrep $\bar{E}_1$ in WP $2c$, \textit{i.e.} for states related to $p$-orbitals.
\end{itemize}

As it can be seen in the basis relations, the irrep orbitals are either pure up-spin or down-spin. Thus, we can identify up to four hopping terms that mix up and down orientations, i.e., that can be interpreted as SOC terms. It turns out that the size of the gap is controlled mostly by two parameters which couple the $\bar E_1 \oplus \bar E_2$ states at WP $2a$ ($d$-type) to the $\bar E_1$ states at $2c$ ($p_z$-type). This observation supports the idea of the gap arising due to hybridisation between the two kinds of orbitals.\newline

To write the Hamiltonian matrix, we order the irrep basis as follows

\begin{equation}
    \{\bar E_2 @ 2a, \bar E_1 @ 2a, \bar E_1 @ 2c\}
\end{equation}

with a total of $2\times2 + 2\times2 + 2\times2$ orbitals and where, for example, $\bar E_2 @ 2a$ means placing an set of orbitals at WP $2a$ transforming as the $\bar E_2$ irrep of the site-symmetry group. For reasons of presentation, we show the Hamiltonian in momentum space, where the Fourier transform convention we chose is

\begin{equation}
    \phi^j(\vec k) = \sum_R e^{i\vec k\cdot(\vec R + \vec r_j)}\phi(\vec R + \vec r_j)
\end{equation}

Here $\vec R$ runs over all the vectors of the direct lattice, and $\vec r_j$ is the position of the orbital in the unit cell.
Under these premises, the Hamiltonian reads 

\begin{widetext}
\begin{equation}\label{eq:full_hamiltonian}
    \left(
\begin{array}{cccccccccccc}
 \epsilon_a & 0 & f_1 & 0 & 0 & 0 & 0 & 0 & f_2^+(k_x) & 0 & f_2^{+*}(k_y) & 0 \\
 0 & \epsilon_a & 0 & f_1 & 0 & 0 & 0 & 0 & 0 & f_2^-(k_x) & 0 & f_2^{-*}(k_y) \\
 0 & 0 & \epsilon_a & 0 & 0 & 0 & 0 & 0 & f_2^-(k_y) & 0 & f_2^{-*}(k_x) & 0 \\
 0 & 0 & 0 & \epsilon_a & 0 & 0 & 0 & 0 & 0 & f_2^+(k_y) & 0 & f_2^{+*}(k_x) \\
 0 & 0 & 0 & 0 & \epsilon_b & 0 & f_1' & 0 & f_3(k_x) & 0 & f_2^*(k_y) & 0 \\
 0 & 0 & 0 & 0 & 0 & \epsilon_b & 0 & f_1' & 0 & f_3(k_x) & 0 & f_3^*(k_y) \\
 0 & 0 & 0 & 0 & 0 & 0 & \epsilon_b & 0 & f_3(k_y) & 0 & f_3^*(k_x) & 0 \\
 0 & 0 & 0 & 0 & 0 & 0 & 0 & \epsilon_b & 0 & f_3(k_y) & 0 & f_3^*(k_x) \\
 0 & 0 & 0 & 0 & 0 & 0 & 0 & 0 & \epsilon_c & 0 & f_4 & 0 \\
 0 & 0 & 0 & 0 & 0 & 0 & 0 & 0 & 0 & \epsilon_c & 0 & f_4 \\
 0 & 0 & 0 & 0 & 0 & 0 & 0 & 0 & 0 & 0 & \epsilon_c & 0 \\
 0 & 0 & 0 & 0 & 0 & 0 & 0 & 0 & 0 & 0 & 0 & \epsilon_c \\
\end{array}
\right)+h.c.
\end{equation}
\end{widetext}

\begin{gather} 
\epsilon_a = a + l\cos kz \\
\epsilon_b = b + m\cos kz \\
\begin{split}
    \epsilon_c = c + (\cos k_x &+ \cos k_y)(e^{-ik_z}\xi + e^{ik_z}\xi') \\
    &+n\cos k_z 
\end{split} \\
f_1 = \cos\frac{k_x}{2}\cos\frac{k_y}{2} \delta
\end{gather}
\begin{gather}f_1' = \cos\frac{k_x}{2}\cos\frac{k_y}{2} \delta' \\
f_2^{\pm}(s) = \pm ite^{-ik_z(Z-1)}\cos\frac{s}{2} \pm iqe^{-ik_z Z}\cos \frac{s}{2} \\
f_3(s) = t'e^{-ik_z(Z-1)}\cos\frac{s}{2} + q'e^{-ik_z Z}\cos\frac{s}{2} \\
f_4 = \tau e^{2ik_z(Z-1)}\cos\frac{k_x}{2}\cos\frac{k_y}{2}
\end{gather}

In the expressions for the matrix elements, all the parameters are real, $Z$ denotes the position along the third axis of the orbitals in WP $2c$ and $s=k_x,k_y$. The different coupling amplitudes correspond to the following:

\begin{itemize}
    \item $a,b,c$ are onsite energies for the $a_{ij}, b_{ij}, c_{ij}$ orbitals
    \item $l,m,n$ are couplings between orbitals $a,b,c$ in adjacent cells in the $z$ direction
    \item $\xi$ and $\xi'$ are hopping terms between $c_{ij}$ orbitals along lattice vectors $(-1,0,-1),(0,-1,-1), (0,1,-1)$ and $(1,0,-1)$ for $\xi$ and their opposites for $\xi'$
    \item $\delta$ / $\delta'$ are couplings between $a_{1j}$ / $b_{1j}$ and $a_{2j}$ / $b_{2j}$ orbitals separated by lattice vectors $(0,-1,0), (0,0,0), (1,-1,0)$ and $(1,0,0)$
    \item $\tau$ mixes $c_{1j}$ orbitals with $c_{2j}$ separated by lattice vectors $(-1,-1,2), (-1,0,2), (0,-1,2)$ and $(0,0,2)$
    \item $t$ / $t'$ are hybridisation nearest-neighbour terms between $a_{ij}$ / $b_{ij}$ orbitals and $c_{1j}$ separated by $(0,0,-1), (1,0,-1)$ and $c_{2j}$ by $(0,-1,1),(0,0,1)$
\end{itemize}

The fitted parameters in Fig. 4 can be seen in Tables S5, S6, and S7.

\begin{table*}[h]
\begin{center}
\caption{TABLE S5: Parameters of the Hamiltonian in Fig. 4(a).}
\begin{tabular}{cccccc}
\hline\hline
\multicolumn{6}{c}{Parameters of fit}                                                        \\ \hline
$a$  & \multicolumn{1}{c|}{-2.33} & $\delta$  & \multicolumn{1}{c|}{5.30}   & $n$   & 1.625   \\ \hline
$b$  & \multicolumn{1}{c|}{-4.40} & $\delta'$ & \multicolumn{1}{c|}{9.25}   & $\xi$ & -1.0  \\ \hline
$c$  & \multicolumn{1}{c|}{-6.13}  & $\tau$    & \multicolumn{1}{c|}{-14.0}  & $q$   & 0.4   \\ \hline
$t$  & \multicolumn{1}{c|}{0.16}  & $l$       & \multicolumn{1}{c|}{-0.049} & $q'$  & -0.01 \\ \hline
$t'$ & \multicolumn{1}{c|}{0.19}   & $m$       & \multicolumn{1}{c|}{-0.049} & -     & -     \\ \hline\hline
\end{tabular}
\label{tab:fit_parameters_1}
\end{center}
\end{table*}

\begin{table*}[h]
\begin{center}
\caption{TABLE S6: Parameters of the Hamiltonian in Fig. 4(b).}
\begin{tabular}{cccccc}
\hline\hline
\multicolumn{6}{c}{Parameters of fit}                                                        \\ \hline
$a$  & \multicolumn{1}{c|}{-2.33} & $\delta$  & \multicolumn{1}{c|}{5.30}   & $n$   & 1.625   \\ \hline
$b$  & \multicolumn{1}{c|}{-4.40} & $\delta'$ & \multicolumn{1}{c|}{9.25}   & $\xi$ & -1.0  \\ \hline
$c$  & \multicolumn{1}{c|}{-6.13}  & $\tau$    & \multicolumn{1}{c|}{-14.0}  & $q$   & 0.4   \\ \hline
$t$  & \multicolumn{1}{c|}{0.16}  & $l$       & \multicolumn{1}{c|}{-0.0163} & $q'$  & -0.01 \\ \hline
$t'$ & \multicolumn{1}{c|}{0.19}   & $m$       & \multicolumn{1}{c|}{-0.0163} & -     & -     \\ \hline\hline
\end{tabular}
\label{tab:fit_parameters_2}
\end{center}
\end{table*}

\begin{table*}[h]
\begin{center}
\caption{TABLE S7: Parameters of the Hamiltonian in Fig. 4(c).}
\begin{tabular}{cccccc}
\hline\hline
\multicolumn{6}{c}{Parameters of fit}                                                        \\ \hline
$a$  & \multicolumn{1}{c|}{-2.33} & $\delta$  & \multicolumn{1}{c|}{5.30}   & $n$   & 1.625   \\ \hline
$b$  & \multicolumn{1}{c|}{-4.40} & $\delta'$ & \multicolumn{1}{c|}{9.25}   & $\xi$ & -1.0  \\ \hline
$c$  & \multicolumn{1}{c|}{-6.13}  & $\tau$    & \multicolumn{1}{c|}{-14.0}  & $q$   & 0.4   \\ \hline
$t$  & \multicolumn{1}{c|}{0.16}  & $l$       & \multicolumn{1}{c|}{-0.0163} & $q'$  & -0.07 \\ \hline
$t'$ & \multicolumn{1}{c|}{0.19}   & $m$       & \multicolumn{1}{c|}{-0.0163} & -     & -     \\ \hline\hline
\end{tabular}
\label{tab:fit_parameters_3}
\end{center}
\end{table*}